\documentclass[11pt,a4paper]{article}
\usepackage{jheppub}

\usepackage{epsfig}
\usepackage{amssymb,amsmath}
\usepackage{pstricks}
\usepackage{pst-all}
\usepackage[utf8]{inputenc}
\usepackage{epstopdf}

\newcommand{\eqa}{\begin{eqnarray}}
\newcommand{\qea}{\end{eqnarray}}
\newcommand{\eq}{\begin{equation}}
\newcommand{\qe}{\end{equation}}

\newcommand{\de}{\mathrm{d}\hspace{0.08em}}

\newcommand{\tr}{\mathrm{Tr}}

\newcommand{\Real}{{\rm Re}}
\newcommand{\avg}[1]{\langle #1 \rangle}
\newcommand{\bigavg}[1]{\Big\langle #1 \Big\rangle}

\title{Centre symmetric 3d effective actions for thermal $SU(N)$
Yang-Mills from strong coupling series}
\author[a]{Jens Langelage,}
\author[b]{Stefano Lottini}
\author[b]{Owe Philipsen}
\affiliation[a]{Fakult\"at f\"ur Physik, Universit\"at Bielefeld, \\
33501 Bielefeld, Germany}
\affiliation[b]{Institut f\"ur Theoretische Physik, Goethe-Universit\"at Frankfurt,\\
Max-von-Laue-Str. 1, 60438 Frankfurt am Main, Germany}
\emailAdd{jlang@physik.uni-bielefeld.de}
\emailAdd{lottini@th.physik.uni-franfurt.de}
\emailAdd{philipsen@th.physik.uni-franfurt.de}
\abstract{We derive three-dimensional, $Z(N)$-symmetric effective actions in terms of Polyakov loops 
by means of strong coupling expansions, starting from thermal $SU(N)$ Yang-Mills theory in four
dimensions on the lattice. An earlier action in the literature, corresponding to the (spatial) strong coupling limit, 
is thus extended by several higher orders, as
well as by additional interaction terms.
We provide analytic mappings between the couplings of the effective theory
and the parameters $N_\tau,\beta$ of the original thermal lattice theory, which can be systematically
improved. We then investigate the deconfinement transition for the cases 
$SU(2)$ and $SU(3)$ by means of Monte Carlo simulations of the effective theory.
Our effective models correctly reproduce second order 3d Ising and first order phase transitions, respectively.
Furthermore, we calculate the critical couplings $\beta_c(N_\tau)$ and find agreement with results from
 simulations of the 4d theory at the few percent level for $N_\tau=4-16$.}
\arxivnumber{1010.0951}

\dedicated{BI-TP-2010/32}

\begin{document}
\maketitle

\section{Introduction}

Non-abelian gauge theories at finite temperature are inherently non-perturbative, due to infrared
problems for soft gauge fields \cite{linde}.
The dynamical appearance of different scales, 
$\pi T, gT, g^2T$ with gauge coupling $g$, has motivated 
effective theory methods, most notably 
dimensional reduction \cite{dr1,dr2}. The idea is to integrate over
the hard scales perturbatively, whereupon a three-dimensional effective theory
arises which is then easier to solve by non-perturbative means. A particularly successful
application is the investigation of the electroweak phase transition as a function of the Higgs mass 
using the 3d effective theories \cite{ew3d1,ew3d2}, which produced results at the few percent accuracy level
compared to simulations of the full 4d theory \cite{ew4d}. 

In QCD, the same technique can be applied to 
observables 
in the deconfined phase only, because the reduction loses validity as the deconfinement transition
is approached. 
An investigation of the phase diagram by similar methods would be particularly desirable,
since lattice Monte Carlo studies at finite baryon density are
beset by the sign-problem. Unfortunately, the QCD phase structure is not inherited
by the dimensionally reduced model, which loses the $Z(N)$ symmetry of Yang-Mills theory
in the perturbative reduction step \cite{drqcd}. This has motivated alternative approaches, where 
the most general $Z(N)$-symmetric theory is written down and the couplings are then
matched to those of the full theory by calculations of particular observables \cite{vy,rob}.
While a succesful description of the phase transition for $SU(2)$ can be achieved in this way \cite{fk}
(see also \cite{dumitru_smith}), in the case of $SU(3)$ there
remain so far several open couplings that cannot be matched easily.

In this paper,
we study the possibility of reducing the full theory to an effective 
theory by using strong coupling expansions on a lattice. 
Such an approach 
was conjectured to be sensible in 
\cite{Svetitsky:1982gs} and has been explored earlier in the literature 
\cite{Polonyi:1982wz,Green:1983sd,Gocksch:1984yk,Gross:1984wb,Heinzl:2005xv,Wozar:2007tz}.
It results in a three-dimensional effective theory of Polyakov loops.
A common simplification was the neglect 
of spatial plaquettes, which has been argued not to influence the universal 
behaviour of the theory. Work where authors went beyond the spatial 
strong coupling limit is \cite{Billo:1996wv}, see also \cite{Nakano:2010bg} 
for recent developments on Polyakov loop extended strong coupling lattice
QCD with staggered fermions.  

Here we significantly extend this approach and calculate longer 
strong coupling series for the effective couplings, which are thus valid beyond 
the spatial strong coupling limit.  
Strong coupling series
have a finite radius of convergence, which in this case is given by the deconfinement transition.
Hence, our effective theory is valid in the confined phase and complementary to weak coupling
approaches.
We also investigate the influence of next-to-nearest neighbour
Polyakov loop interaction terms which arise in higher orders. 
For our first studies we consider the pure gauge 
theories $SU(2)$ and $SU(3)$. After derivation of the effective theories of complex scalar
fields in 3d, corresponding to the traces of the Polyakov loop and its hermitian conjugate,
we study them by means of Monte Carlo simulations. In particular we determine the
order and the critical couplings of the deconfinement phase transition and relate them to those of 
the original theories. We find qualitative and quantitative agreement with less than 6\% deviations
for the $\beta_c(N_\tau)$ up to $N_\tau=16$, suggesting that these models are useful for
the study of continuum physics.

The paper is structured as follows: in section \ref{sec:deriv} the effective actions for the two 
gauge groups are derived, with detailed discussion about the higher-order contributions 
to the effective couplings; then, in section \ref{sec:numerical}, we present the Monte Carlo 
implementation of the effective models together with numerical results; section \ref{sec:results} 
provides the corresponding predictions for the original 4d theories and in section \ref{sec:conclusions} 
we conclude.

\section{Derivation of the effective theory}
\label{sec:deriv}

\subsection{General strategy and $SU(2)$}

Consider the partition function of a $(3+1)$-dimensional lattice gauge
field theory at finite temperature $\left(T=\frac{1}{aN_\tau}\right)$ with 
gauge group $SU(N)$ and Wilson's gauge action
\begin{eqnarray}
Z=\int\left[dU_0\right]\left[dU_i\right]\exp\left[\frac{\beta}{2N}\sum_p
\left(\mathrm{tr}\;U_p+\mathrm{tr}\;U_p^{\dagger}\right)\right], \quad \beta=\frac{2N}{g^2}\;.
\label{eq:original_gaugetheory}
\end{eqnarray}
Finite temperature and the bosonic nature of the degrees of freedom imply the use of 
periodic boundary conditions in the time direction.

In order to arrive at an effective three-dimensional theory, we integrate
out the spatial degrees of freedom and get schematically
\cite{Gross:1984wb}:
\begin{eqnarray}
Z&=&\int\left[dU_0\right]\exp\left[-S_\mathrm{eff}\right]\;;\nonumber\\
-S_\mathrm{eff}&=&\ln\int\left[dU_i\right]\exp\left[\frac{\beta}{2N}\sum_p
\left(\mathrm{tr}\;U_p+\mathrm{tr}\;U_p^{\dagger}\right)\right]\;\equiv\nonumber\\
&\equiv&\lambda_1S_1+\lambda_2S_2+\ldots\;.
\label{eq_seff}
\end{eqnarray}
We expand around $\beta=0$ and arrange the
effective couplings $\lambda_n=\lambda_n(\beta,N_\tau)$ 
in increasing order in $\beta$ of their leading terms.
Thus, the $\lambda_n$ become less 
important the higher $n$. 
As we shall see, the interaction terms
$S_n$ depend only on Polyakov loops
\begin{eqnarray}
 L_j\equiv\mathrm{tr}\;W_j\equiv
\mathrm{tr}\;\prod_{\tau=1}^{N_\tau}U_0(\vec{x}_j,\tau)\;.
\end{eqnarray}
This is the reason for a ``dimensional reduction'' occurring here, as the time dimension
is now implicit in the variables of the effective theory, which are fields defined on 
the spatial lattice.
With sufficiently accurate knowledge of the relations 
$\lambda_n(\beta,N_\tau)$, we are able to convert the 
couplings of the three-dimensional theory to those of the full theory.
In this work we are mainly interested in the deconfinement transition.  Determining the 
critical parameters $\lambda_{n,c}$ 
of the effective theory then gives a whole
array of critical $\beta_c(N_\tau)$ for - in principle - all $N_\tau$. 
In the following we calculate strong coupling, i.e.~small $\beta$, expansions  
of the leading $\lambda_{n}$.

Since the calculations are quite similar for different numbers of colours, we now specialise
our derivation to the simpler case of $SU(2)$ and later provide the necessary changes for $SU(3)$. 
Using the character expansion as described e.g.~in \cite{Montvay:1994cy,
Drouffe:1983fv}, the 
effective action according to Eq.~(\ref{eq_seff}) can be written as
\begin{eqnarray}
-S_\mathrm{eff}=\ln\int\left[dU_i\right]\prod_p\left[1+\sum_{r\neq0}d_ra_r(\beta)
\chi_r(U_p)\right]\;,
\label{eq_char}
\end{eqnarray}
where the sum extends over all irreducible representations $r$ with dimension 
$d_r$ and character $\chi_r$. The expansion coefficients $a_r(\beta)$ are 
accurately known 
\cite{Montvay:1994cy}
and in the following we use $u\equiv a_f$  
as expansion parameter instead of $\beta$ for its better apparent convergence.
The logarithm in this definition allows us to use the method of moments
and cumulants \cite{Munster:1980wc}, and we get the following cluster expansion 
\begin{eqnarray}
-S_\mathrm{eff}&=&\sum_{C=(X_l^{n_l})}a(C)\prod_l\Phi\Big(X_l;\left\lbrace 
W_j\right\rbrace\Big)^{n_l}\;;\\
\Phi\Big(X_l;\left\lbrace 
W_j\right\rbrace\Big)&=&\int\left[dU_i\right]\prod_{p\in X_l}d_{r_p}a_{r_p}
\chi_{r_p}(U_p)\;,\nonumber
\label{eq_charexp}
\end{eqnarray}
where the combinatorial factor $a(C)$ equals 1 for a single polymer $X_i$ 
and $-1$ for two non-identical connected polymers. For clusters consisting of more than two 
polymers, $a(C)$ depends on how these polymers are connected. In contrast to 
earlier calculations \cite{Langelage:2008dj,Langelage:2009jb}, where also the temporal
links were integrated over, the contribution 
$\Phi$ still depends on Wilson line variables $W_j$, which results in the 
interaction terms for the effective action, Eq.~(\ref{eq_seff}). Our task is then to
group together all graphs yielding the same interaction terms up to some order 
in $\beta$, and this finally gives the strong coupling expansion 
of the corresponding effective coupling $\lambda_n$.

\subsection{The spatial strong coupling limit}

In earlier strong coupling calculations at finite temperature it has been
customary to neglect spatial plaquettes. Let us briefly discuss
this limit as it gives some important insights for the following. Neglecting
spatial plaquettes, 
the integrations in Eq.~(\ref{eq_char}) can be done  
applying the group integration rules
\begin{eqnarray}
\int dU\chi_r(XU)\chi_s(U^{-1}Y)&=&\frac{\delta_{rs}}{d_r}\chi_r(XY)\\
\longrightarrow\qquad\int dU\chi_r(U)&=&\delta_{r,0}
\end{eqnarray}
on each spatial link
and we get the partition function \cite{Green:1983sd}:
\begin{eqnarray}
Z=\int\left[dW\right]\prod_{<ij>}\left[1+\sum_{r\neq0}
\Big[a_r(\beta)\Big]^{N_\tau}
\chi_r(W_i)\chi_r(W_j)\right]\;,
\label{eq_sscl}
\end{eqnarray}
where we replaced the integration over all
temporal links with an integration over the Wilson line variables 
$W_i$ as in \cite{Mathur:1995gr}. This is justified since the integrand 
depends on temporal link variables only through the $W_i$.

Some important observations in this limit are:
\begin{itemize}
\item The summation extends only over pairs of nearest neighbours $<$$ij$$>$, 
i.e.~next-to-nearest neighbour interactions vanish without the inclusion of
spatial plaquettes.
\item Contributions from higher representations start with higher powers of $\beta$.
\item The exponential function has cancelled, hence if we insist on having
a partition function of the form $Z=\int\exp[-S]$, we have to introduce
a logarithm in the action.
\end{itemize}

\subsection{Leading order effective action}

\begin{figure}
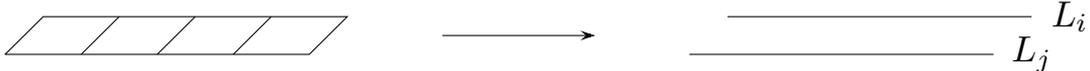

\hspace{-3.5cm}
\scalebox{0.5}{
\psline(8,0)(16,0)(17,1)(9,1)(8,0)
\psline(10,0)(11,1)
\psline(12,0)(13,1)
\psline(14,0)(15,1)
\psline(19.5,0.5)(23.5,0.5)
\psline[linewidth=3pt]{->}(23.4,0.5)(23.5,0.5)
\psline(26,0)(34,0)%\psdot[dotscale=2 2](20,0)\psdot[dotscale=2 2](28,0)
\psline(27,1)(35,1)%\psdot[dotscale=2 2](21,1)\psdot[dotscale=2 2](29,1)
\rput(36,1){\scalebox{2.5}{$L_i$}}
\rput(35,0){\scalebox{2.5}{$L_j$}}
}
\caption{First graph with a nontrivial contribution 
after spatial integration for a lattice with temporal extent 
$N_\tau=4$. Four plaquettes in the fundamental representation lead to an 
interaction term involving two adjacent fundamental Polyakov loops 
 $L_i$ and $L_j$.}
\label{fig_lo}
\end{figure}
The leading order result of the effective action has first been 
calculated in \cite{Polonyi:1982wz} and corresponds to a sequence
of $N_\tau$ plaquettes that wind around the lattice in temporal direction, 
cf.~figure \ref{fig_lo}.
Its contribution is given by:
\begin{eqnarray}
\lambda_1S_1=u^{N_\tau}\sum_{<ij>}L_iL_j\;.
\end{eqnarray}
Hence, to leading order the first coupling of the effective theory is 
$\lambda_1(u,N_\tau)=u^{N_\tau}$.

\begin{figure}[t!]
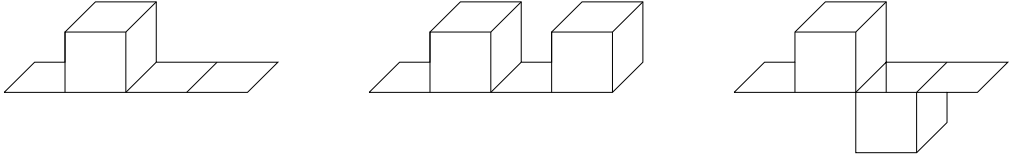

%\vspace{1.5cm}
\hspace{1cm}
\scalebox{0.4}{
\psline(0,0)(8,0)(9,1)(5,1)(5,3)(3,3)(2,2)(2,1)(1,1)(0,0)
\psline(2,0)(2,2)(4,2)(5,3)
\psline(4,0)(4,2)
\psline(4,0)(5,1)
\psline(6,0)(7,1)
\psline(12,0)(20,0)(21,1)(21,3)(19,3)(18,2)(18,1)(17,1)(17,3)(15,3)(14,2)(14,1)(13,1)(12,0)
\psline(14,0)(14,2)(16,2)(17,3)
\psline(16,0)(16,2)
\psline(16,0)(17,1)
\psline(18,0)(18,2)(20,2)(21,3)
\psline(20,0)(20,2)
\psline(24,0)(28,0)(28,-2)(30,-2)(31,-1)(31,0)(32,0)(33,1)(29,1)(29,3)(27,3)
(26,2)(26,1)(25,1)(24,0)
\psline(26,0)(26,1)
\psline(26,2)(28,2)(29,3)
\psline(28,0)(28,2)
\psline(28,0)(29,1)(29,0)
\psline(30,-2)(30,0)
\psline(28,0)(31,0)
\psline(30,0)(31,1)
}
\vspace{0.5cm}
\caption{Left: First correction to the leading order graph, proportional 
to $N_\tau$. Middle, right: Repetitions of this decoration.}
\label{fig_exp}
\end{figure}

For additional terms of the series for $\lambda_1$, we 
can use most of the graphs that also appear in the strong coupling expansion of the 
Polyakov loop susceptibility
\cite{Langelage:2009jb}. These corrections 
involve additional plaquettes, are hence of higher order in $u$ and we 
call these attached plaquettes decorations. Let us note that repetitions of 
lower order decorations attached to planar graphs exponentiate and we can 
write
\begin{eqnarray}
\lambda_1(u,N_\tau)=u^{N_\tau}\exp\Big[N_\tau P(u,N_\tau)\Big]\;,
\end{eqnarray}
with some polynomial $P(u,N_\tau)$. This can be seen e.g.~from the graphs shown 
in figure \ref{fig_exp} 
and their corresponding contributions:
\begin{eqnarray}
\mbox{Left:}\qquad\Phi_1&=&u^{N_\tau}\Big[4N_\tau u^4\Big]S_1\;;\nonumber\\
\mbox{Middle:}\qquad\Phi_2&=&u^{N_\tau}\bigg[\frac{1}{2!}
\left(4N_\tau u^4\right)\cdot4(N_\tau-3)u^4\bigg]S_1\;;\nonumber\\
\mbox{Right:}\qquad\Phi_3&=&u^{N_\tau}\bigg[4N_\tau u^4
\cdot 3N_\tau u^4\bigg]S_1\;.
\end{eqnarray}
Combining the three parts, we can write this, up to higher orders, as
\begin{eqnarray}
\Phi_1+\Phi_2+\Phi_3&=&u^{N_\tau}\exp\Big[N_\tau\left(4u^4-12u^8\right)\Big]S_1\;.
\end{eqnarray}
Of course, the polynomial in the exponential is only part of the complete 
result for $\lambda_1$ to that order. For example,  
there are other graphs contributing to order $u^6$ which are still missing in this correction. 
Nevertheless, in this exponentiated form the effective coupling 
corresponds to a partial resummation of higher order 
terms which may be expected to improve convergence behaviour.
% Let us remark that also in \cite{Munster:1980wc} such an exponentiation 
% has been observed for the strong coupling expansion of the string tension. 
Let us remark that such an exponentiation 
has been observed also for the strong coupling expansion of the string tension. 

\begin{figure}[t!]
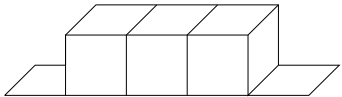

\hspace{5cm}\scalebox{0.4}{
\psline(0,0)(10,0)(11,1)(9,1)(9,3)(3,3)(2,2)(2,1)(1,1)(0,0)
\psline(2,0)(2,2)(8,2)(9,3)
\psline(4,0)(4,2)(5,3)
\psline(6,0)(6,2)(7,3)
\psline(9,1)(8,0)(8,2)
}
\vspace{0.5cm}
\caption{Shape of the most elongated graph contributing to $\lambda_1$ at order $n=2(q+1)$, here in the case $q=3$:
only systems with $N_\tau>q$ can actually accommodate for it.}
\label{fig:elongated_graph}
\end{figure}

Carrying out the calculations, we get the following results through
order $u^{10}$ 
in the corrections relative to the leading order graph:
\begin{eqnarray}
\lambda_1(u,2)&=&u^2\exp\left[2\left(4u^4-8u^6+\frac{134}{3}u^8-\frac{49044}{405}u^{10}\right)\right]\;,\nonumber\\
\lambda_1(u,3)&=&u^3\exp\left[3\left(4u^4-4u^6+\frac{128}{3}u^8-\frac{36044}{405}u^{10}\right)\right]\;,\nonumber\\
\lambda_1(u,4)&=&u^4\exp\left[4\left(4u^4-4u^6+\frac{140}{3}u^8-\frac{37664}{405}u^{10}\right)\right]\;,\nonumber\\
\lambda_1(u,N_{\tau}\geq5)&=&u^{N_\tau}\exp\left[N_{\tau}\left(4u^4-4u^6+\frac{140}{3}u^8-\frac{36044}{405}u^{10}\right)\right]\;.
\label{eq_lambda}
\end{eqnarray}
For smaller $N_\tau$ some graphs do not contribute since the temporal 
extent of their 
decoration is $\geq N_\tau$ so that they do not fit into the lattice.
The coefficients of the order $n$ in the exponents 
of the effective couplings, Eq.~(\ref{eq_lambda}), reach
their $N_\tau=\infty$ values
as soon as $N_\tau=n/2$. This can be understood as follows. Among all graphs contributing to the coefficient at order $u^n$, 
the most elongated in the time direction is of the type depicted in figure \ref{fig:elongated_graph}, with $q$
consecutive lifted plaquettes, which brings in $n=2(q+1)$ additional 
powers of $u$. Such a graph
is included only if $N_\tau\geq q+1$ (in the 
moment-cumulant formalism a given polymer cannot occupy twice 
the same plaquette).

\subsection{Higher order terms}

There occur several types of higher order graphs: larger numbers of loops involved, Polyakov 
loops at distances larger than one and Polyakov loops in 
higher dimensional representations.

We begin by considering powers of the leading order term.
Inspection of higher order terms shows that one can arrange a subclass of 
these terms in the following manner
\begin{eqnarray}
\sum_{<ij>}\left(\lambda_1L_iL_j-\frac{\lambda_1^2}{2}L_i^2L_j^2+
\frac{\lambda_1^3}{3}L_i^3L_j^3-\ldots\right)=\sum_{<ij>}\ln\left(1+
\lambda_1L_iL_j\right)\;.
\label{eq_powers}
\end{eqnarray}
Thus, there are graphs that reproduce the emergence of the logarithm just as in 
the spatial strong coupling result, Eq.~(\ref{eq_sscl}). In contrast 
to that case we have now the full effective coupling $\lambda_1(u,N_\tau)$ appearing in the
logarithm instead of only 
its leading order term $u^{N_\tau}$, which results if we restrict 
Eq.~(\ref{eq_sscl}) to the fundamental representation. To see this, one 
calculates the corresponding graphs with $L_i^2L_j^2$ or $L_i^3L_j^3$, and the 
combinatorial factor $a(C)$ of Eq.~(\ref{eq_charexp}) gives the correct prefactors 
for the series to represent a logarithm. 

\begin{figure}[t!]
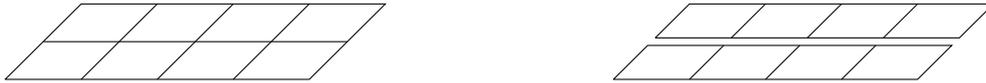

\hspace{1cm}
\scalebox{0.5}{
\psline(0,0)(8,0)(10,2)(2,2)(0,0)
\psline(1,1)(9,1)
\psline(2,0)(4,2)
\psline(4,0)(6,2)
\psline(6,0)(8,2)
\psline(16,0)(24,0)(24.9,0.9)(16.9,0.9)(16,0)
\psline(18,0)(18.9,0.9)
\psline(20,0)(20.9,0.9)
\psline(22,0)(22.9,0.9)
\psline(17.1,1.1)(25.1,1.1)(26,2)(18,2)(17.1,1.1)
\psline(19.1,1.1)(20,2)
\psline(21.1,1.1)(22,2)
\psline(23.1,1.1)(24,2)
}
\caption{Two graphs without spatial plaquettes that cancel against each other 
due to different signs.}
\label{fig_cancel}
\end{figure}

\begin{figure}[t!]
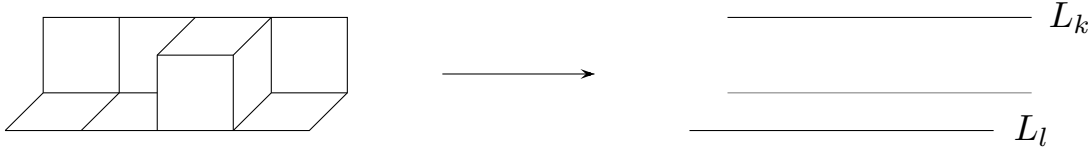

\vspace{1.5cm}
\hspace{0.5cm}
\scalebox{0.5}{
\psline(0,0)(8,0)(9,1)(9,3)(1,3)(1,1)(0,0)
\psline(2,0)(3,1)(3,3)
\psline(4,0)(4,2)(5,3)
\psline(6,0)(6,2)(7,3)
\psline(6,0)(7,1)(7,3)
\psline(1,1)(4,1)
\psline(4,2)(6,2)
\psline(7,1)(9,1)
\psline(11.5,1.5)(15.5,1.5)
\psline[linewidth=3pt]{->}(15.4,1.5)(15.5,1.5)
\psline(18,0)(26,0)
\psline[linecolor=gray](19,1)(27,1)
\psline(19,3)(27,3)
%\psline[arrowsize=0.3]{->}(21.9,0)(22,0)
%\psline[arrowsize=0.3]{->}(23.1,3)(23,3)
\rput(28,3){\scalebox{2.5}{$L_k$}}
\rput(27,0){\scalebox{2.5}{$L_l$}}
}
\caption{Leading order graph contributing to the interaction between loops at a distance $\sqrt{2} a$, 
	denoted with $[kl]$ in the summations.}
\label{fig:lshaped-sketch}
\end{figure}

\begin{figure}[t!]
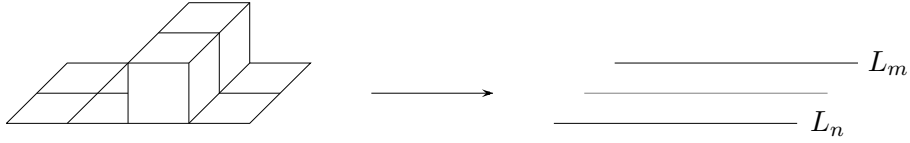

\hspace{-1.5cm}
\scalebox{0.4}{
\psline(8,0)(16,0)(18,2)(16,2)(16,4)(14,4)(12,2)(10,2)(8,0)
\psline(10,0)(12,2)
\psline(9,1)(12,1)
\psline(12,0)(12,2)
\psline(13,3)(15,3)(15,1)(17,1)
\psline(12,2)(14,2)
\psline(14,0)(14,2)(16,4)
\psline(14,0)(16,2)
\psline(20,1)(24,1)
\psline[linewidth=3pt]{->}(23.9,1)(24,1)
\psline(26,0)(34,0)
\psline(28,2)(36,2)
\psline[linecolor=gray](27,1)(35,1)
\rput(37,2){\scalebox{2.5}{$L_m$}}
\rput(35,0){\scalebox{2.5}{$L_n$}}
}
\caption{Leading order graph of next-to-nearest neighbours with distance $2a$.}
\label{fig_nlo}
\end{figure}

Next, let us consider couplings pertaining
to next-to-nearest neighbour interactions. These appear once additional
plaquettes are taken into account. Naively, the leading contribution should 
correspond to a planar graph with Polyakov loops at distance two. However, this
graph is precisely cancelled by the contribution of the nearest-neighbour graph squared
and its associated combinatorial factor $-1$ (figure \ref{fig_cancel}).
The leading non-zero contribution therefore comes from L-shaped graphs and is given by
\begin{eqnarray}
\lambda_2(u,N_\tau)S_2=N_\tau(N_\tau-1)u^{2N_\tau+2}\sum_{\left[kl\right]}L_kL_l\;,
\end{eqnarray}
where we have two additional spatial plaquettes (figure \ref{fig:lshaped-sketch}) and we sum over all pairs 
of loops with a diagonal distance of $\sqrt{2}a$, abbreviated by 
$\left[kl\right]$. With the same steps leading 
to Eq.~(\ref{eq_powers}), we finally arrive at the $SU(2)$ partition 
function
\begin{eqnarray}
Z=\int\left[dW\right]\prod_{<ij>}\left[1+\lambda_1L_iL_j\right]\prod_{\
\left[kl\right]}\left[1+\lambda_2L_kL_l\right]\;.
\label{eq:2coupling_su2}
\end{eqnarray}
Let us further point out that corrections to the planar next-to-nearest neighbours, 
having a distance of $2a$, give to leading order
\begin{eqnarray}
\lambda_3(u,N_\tau)S_3 \sim u^{2N_\tau+6}\sum_{<<mn>>}L_mL_n\;,
\end{eqnarray}
with an obvious notation for straight line next-to-nearest neighbours. The 
leading order graph is given in figure \ref{fig_nlo}. Since these corrections are of higher 
order in $u$ than the ones with distance $\sqrt{2} a$, we will omit them in this work.
As we shall see later, the effective theory in case of $SU(2)$ works quite well 
even without next-to-nearest neighbours, hence we will quote explicit results for 
the coupling $\lambda_2$ for the case of $SU(3)$ only.

Finally, we include some remarks about the Polyakov loops in higher 
dimensional representations. Consider, e.g., the adjoint Polyakov loop: the 
leading order term 
emerging from a strong coupling expansion is
\begin{eqnarray*}
\lambda_aS_a=v^{N_\tau}\sum_{<ij>}\chi_a(W_i)\chi_a(W_j)\;,\quad v=\frac{2}{3}u^2
+\frac{2}{9}u^4+\frac{16}{135} u^6+\ldots
\end{eqnarray*}
and hence $\lambda_a\sim u^{2N_\tau}$, which is formally of lower order than the coupling
$\lambda_2$. To next-to-leading order (valid for all $N_\tau\geq 2$) we have
\begin{equation}
\lambda_a=v^{N_\tau}\left( 1+N_\tau \frac{8}{3}\frac{u^6}{v}+\ldots \right).
\end{equation} 
Effects of higher representations have also been investigated in the literature 
\cite{Heinzl:2005xv,Wozar:2007tz,Dumitru2003:hp}.

\subsection{The effective action for $SU(3)$}

In the case of $SU(3)$ the same steps as for $SU(2)$ apply.
The only difference we have to keep in mind is that $SU(3)$ also has 
an anti-fundamental representation and consequently there is also
a complex conjugate Polyakov loop variable $L_i^{\ast}$.
Thus we get the one-coupling and two-coupling partition functions
\begin{eqnarray}
Z_1&=&\int\left[dW\right]\prod_{<ij>}\left[1+\lambda_1\left(L_iL_j^{\ast}+
L_i^{\ast}L_j\right)\right]\;,\label{eq:su3-onecoupling}\\
Z_2&=&\int\left[dW\right]\prod_{<ij>}\left[1+\lambda_1\left(L_iL_j^{\ast}+
L_i^{\ast}L_j\right)\right]\prod_{\left[kl\right]}\left[1+\lambda_2\left(L_kL_l^{\ast}+
L_k^{\ast}L_l\right)\right]\;.\label{eq:su3-twocoupling}
\end{eqnarray}
The effective coupling $\lambda_1(u,N_\tau)$ is obtained as 
(for this gauge group, in view of the comparison that will be made with available 4d Yang-Mills simulation data,
 we consider only even values of $N_\tau$; however, analogous formulae hold for odd $N_\tau$):
\begin{eqnarray}
\hspace{-0.4cm}\lambda_1(2,u)&=&u^2\exp\left[2\left(4u^4+12u^5-18u^6-36u^7\right.\right.\nonumber\\
&&\hspace*{2cm} \left.\left.
+\frac{219}{2}u^8+\frac{1791}{10}u^9+\frac{830517}{5120}u^{10}\right)\right]\;,\nonumber\\
\hspace{-0.4cm}\lambda_1(4,u)&=&u^4\exp\left[4\left(4u^4+12u^5-14u^6-36u^7\right. \right. \nonumber\\
&&\hspace*{2cm}\left.\left.
+\frac{295}{2}u^8+\frac{1851}{10}u^9+\frac{1035317}{5120}u^{10}\right)\right]\nonumber\;,\\
\hspace{-0.4cm}\lambda_1(N_{\tau}\geq6,u)&=&u^{N_\tau}\exp\left[N_{\tau}\left(4u^4+12u^5-14u^6-36u^7\right.\right.\nonumber\\
&&\hspace*{2.5cm}\left.\left.
+\frac{295}{2}u^8+\frac{1851}{10}u^9+\frac{1055797}{5120}u^{10}\right)\right]\;.
	\label{eq_lambda2}
\end{eqnarray}

For the first terms of the next-to-nearest neighbour coupling 
$\lambda_2(N_\tau,u)$ we find
\begin{eqnarray}
\lambda_2(2,u)&=&u^4\Big[2u^2+6u^4+31u^6\Big]\;,\nonumber\\
\lambda_2(4,u)&=&u^8\Big[12u^2+26u^4+364u^6\Big]\;,\nonumber\\
\lambda_2(6,u)&=&u^{12}\Big[30u^2+66u^4\Big]\;,\nonumber\\
\lambda_2(N_\tau\geq8,u)&=&u^{2N_\tau}\left[N_\tau(N_\tau-1)u^2\right]\;,
\end{eqnarray}
while the leading coupling of adjoint loops is (valid for $N_\tau\geq 2$)
\begin{equation}
\lambda_a=v^{N_\tau}\left( 1+N_\tau \frac{3}{2}\frac{u^6}{v}+\ldots \right),\quad
v=\frac{9}{8}u^2-\frac{9}{8}u^3+\frac{81}{32}u^4+\ldots
\label{adjlam}
\end{equation}

\section{Numerical simulation of the effective theories}
\label{sec:numerical}

In this section we proceed to evaluate the effective theories 
by Monte Carlo methods. 
After introducing a suitable algorithm,  
we discuss the strategy employed
to extract critical couplings and the properties of phase transitions,
and present detailed numerical results. 

\subsection{The one coupling model}

\begin{figure}
	\begin{center}
	\includegraphics[width=1.8cm]{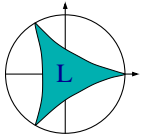}
	\caption[]{Domain for the complex numbers $L_x$ in the $SU(3)$ effective theory.}
	\label{fig:su3_L_domain}
	\end{center}
\end{figure}

For the purpose of numerical simulations, a further simplification is achieved by using 
the trace of the Polyakov loops for the path integral measure as degrees of freedom 
(complex numbers instead of matrices), and rewrite the one-coupling partition function for 
$SU(3)$, Eq.~(\ref{eq:su3-onecoupling}),
\eq
	Z = \Big(\prod_x \int \de L_x\Big) e^{-S_\mathrm{eff}}
		\;;\;
	S_\mathrm{eff} = -\sum_{<ij>} \log(1+2\lambda_1 \Real L_i L^*_j)-
\sum_x V_x\;.
\qe
The complex numbers $L_x$ represent traces of $SU(3)$ 
matrices, and thus $|L_x|\leq 3$.
The complex domain of $L_x$ is thus the 
``three-pointed star'' with radius $3$, figure \ref{fig:su3_L_domain}.

The potential term in $S_\mathrm{eff}$ is the Jacobian induced by the Haar measure of 
the original group integration and its
actual form depends on the chosen parametrisation for $L_x$.
Following \cite{gross_bartholomew_hochberg_1983}, we write the trace of a 
$SU(3)$ matrix by rotating it to its diagonal form,
\eq
	L_x(\theta,\phi) = e^{i\theta}+e^{i\phi}+e^{-i(\theta+\phi)}\;,\;-
\pi \leq \theta,\phi\leq +\pi\;.
\qe
In this case the potential is
\eq
	V_x = \frac{1}{2}\log(27-18|L_x|^2+8\Real L_x^3-|L_x|^4)\;.
\qe
The integration measure actually used in our simulation then takes the form
\eq
	\int \de L_x e^{V_x} = \int_{-\pi}^{+\pi}\de\phi_x
		\int_{-\pi}^{+\pi}\de\theta_x e^{V_x}\;.
\qe
When working on the $SU(2)$ theory, $-2\leq L_x \leq +2$ is a real number 
and we simply have 
\eq
	\int_{-2}^{+2}\de L_x e^{V_x}\;,\;V_x = \frac{1}{2}\log(4-L_x^2)\;.
\qe

\subsection{A ``sign problem'' and its solution}

The effective theory described so far has lower dimensionality and 
simpler degrees of freedom per site than
the initial Wilson action, suggesting a 
straightforward local Metropolis update based
on accept/reject steps. This ideal 
situation is somewhat complicated by some sort of a ``sign problem''.

Consider the effective partition function in
Eq.~(\ref{eq:su3-onecoupling}). For couplings larger than some threshold, 
\mbox{$\lambda > \lambda^T$}, there are
gauge configurations yielding negative contributions to $Z$, or, 
equivalently, for which the logarithms appearing in the effective
action have negative arguments. Although $Z$ remains overall positive, 
the Metropolis update ceases working when approaching this threshold
value of the coupling.  

For the $SU(3)$ action $\Real L_i L^*_j$ can go down to $-9/2$ and the threshold coupling 
is $\lambda^T=1/9$, while for $SU(2)$ we have $\lambda^T = 1/4$. 
In this work our interest is in the phase transition, and we shall see that 
for $SU(3)$ the critical coupling $\lambda_{1,c}$ happens to be
close to $\lambda^T$. This sign problem is 
then a practical concern and a solution is called for.
Fortunately, in the case of $SU(2)$ we have $\lambda_{1,c} \sim 0.2 < \lambda^T = 
1/4$ and this problem can 
be ignored. We exploit this fact to test our  workaround for the problem
against the full solution. 

Our approach to overcome this problem is the following: we Taylor-expand the 
logarithm in the effective action to some order $M$ in powers
of $q \equiv \lambda_1\Real L_i L^*_j$, which effectively undoes 
the resummation as in Eq.~(\ref{eq_powers}),
	\eq
		S_\mathrm{eff}^{(M)} = -\sum_x V_x -\sum_{<ij>} \Big(
			2q -2q^2 +\frac{8}{3}q^3 - 4q^4+\frac{32}{5}q^5 - 
			\ldots - (-1)^M \frac{2^M}{M}q^M\Big)\;.
		\label{eq:taylor-m-expansion}
	\qe
We thus obtain a family of models which are, by construction, free 
of the mentioned problem, and are expected
to converge to the full theory as $M\to\infty$. The strategy 
is then to find the series of critical points $\lambda_{1,c}(M)$ 
and to look for their convergence
behaviour as $M$ increases.

In order to have a critical model, moreover, one must truncate the 
expansion at an odd power $M$: for even truncations, indeed, the slope of
the effective action as a function of $q$, corresponding to a force, vanishes for $2q \lesssim 1$,
and never triggers a symmetry breaking by favouring
aligned neighbours. Let us observe,
incidentally, that this limitation is absent in the $SU(2)$ case, exactly 
because the transition region is far from the
threshold coupling and the value of $\lambda_1 L_iL_j$ is always too far from $1$ for this 
phenomenon to take place.

\subsection{Phase structure, critical coupling and finite size analysis}

\begin{figure}
\hspace*{-0.5cm}
\includegraphics[width=0.25\textwidth,angle=-90]{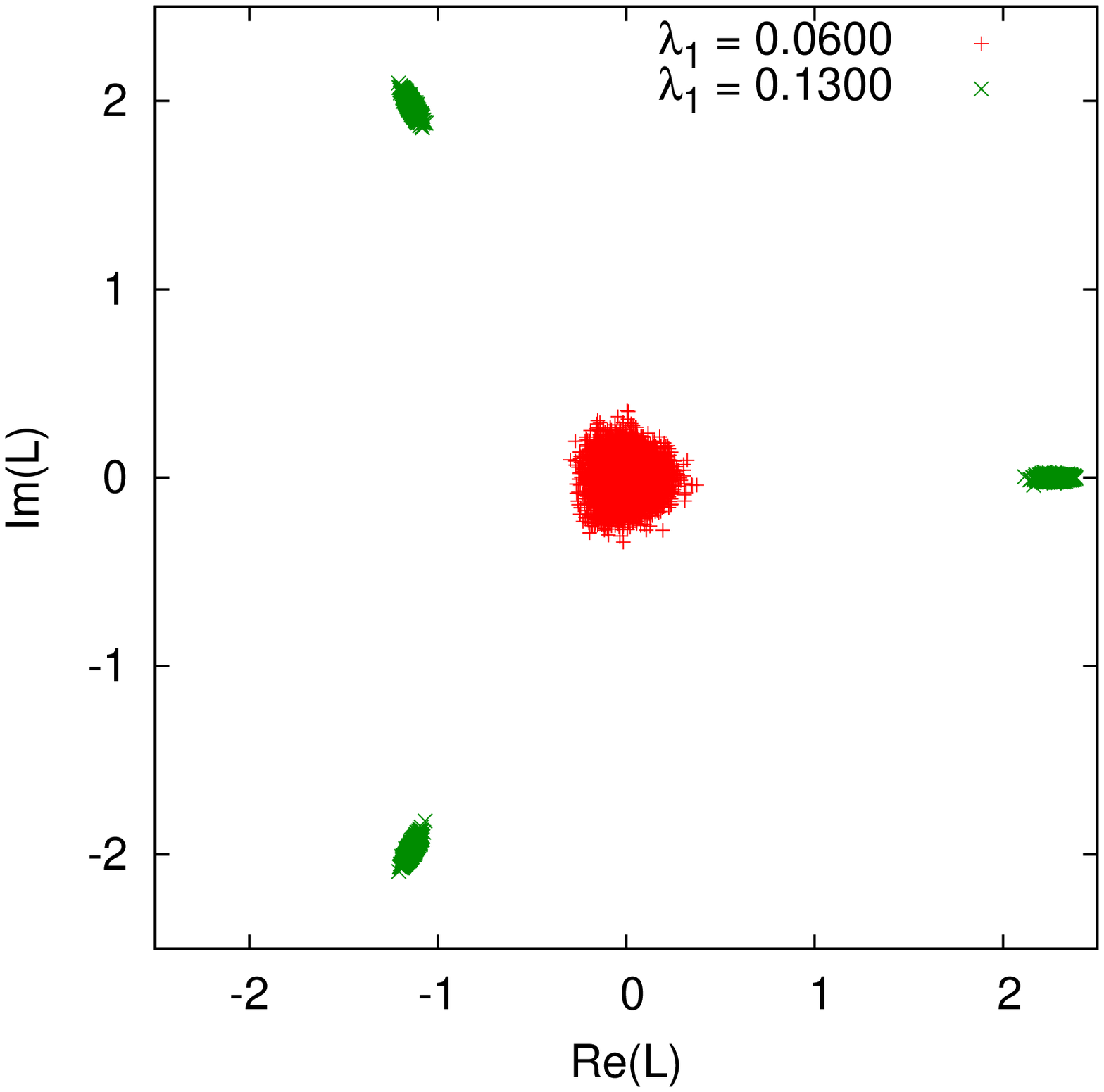}
\hspace*{-0.8cm}
\includegraphics[width=0.25\textwidth,angle=-90]{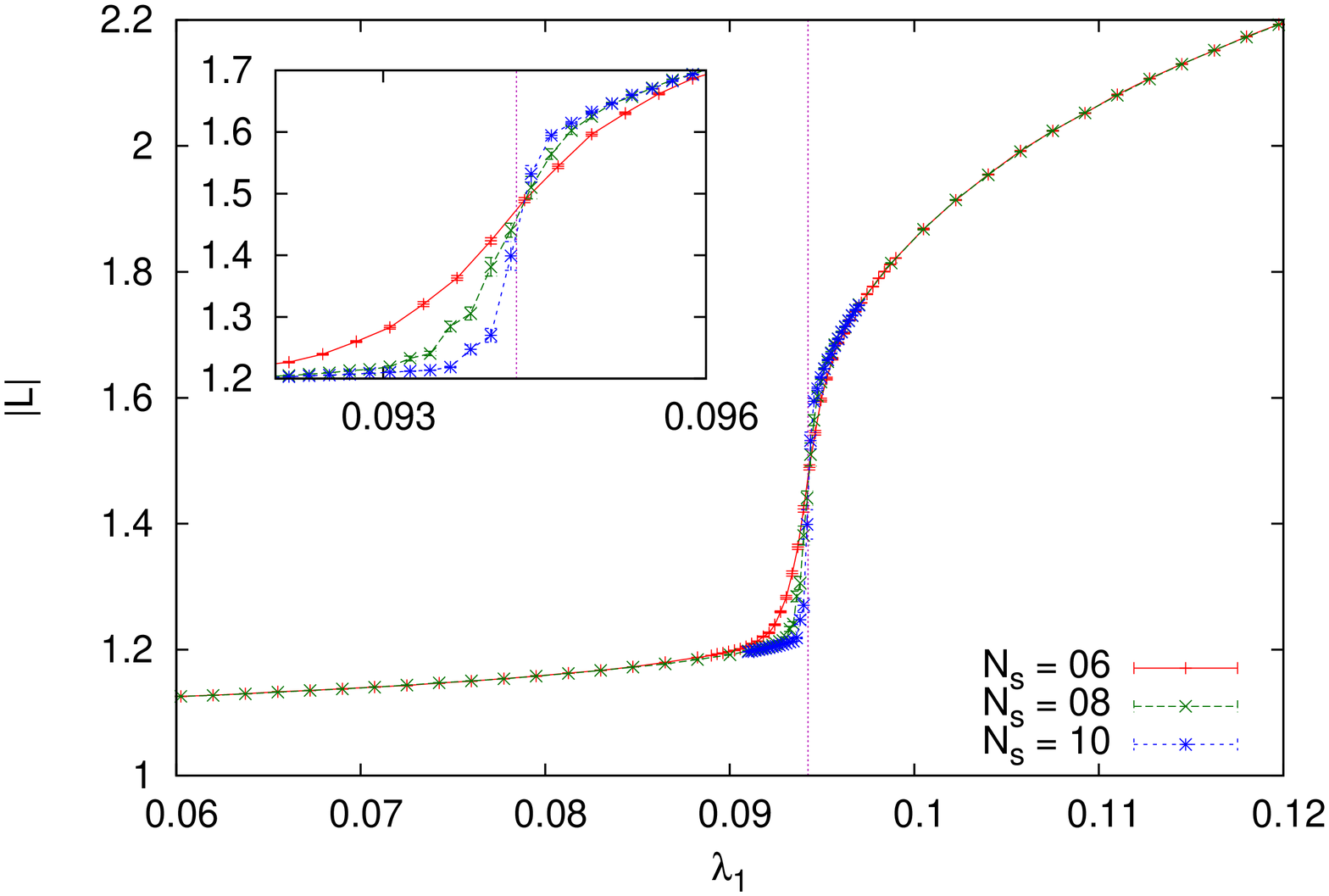}~\includegraphics[width=0.25\textwidth,angle=-90]{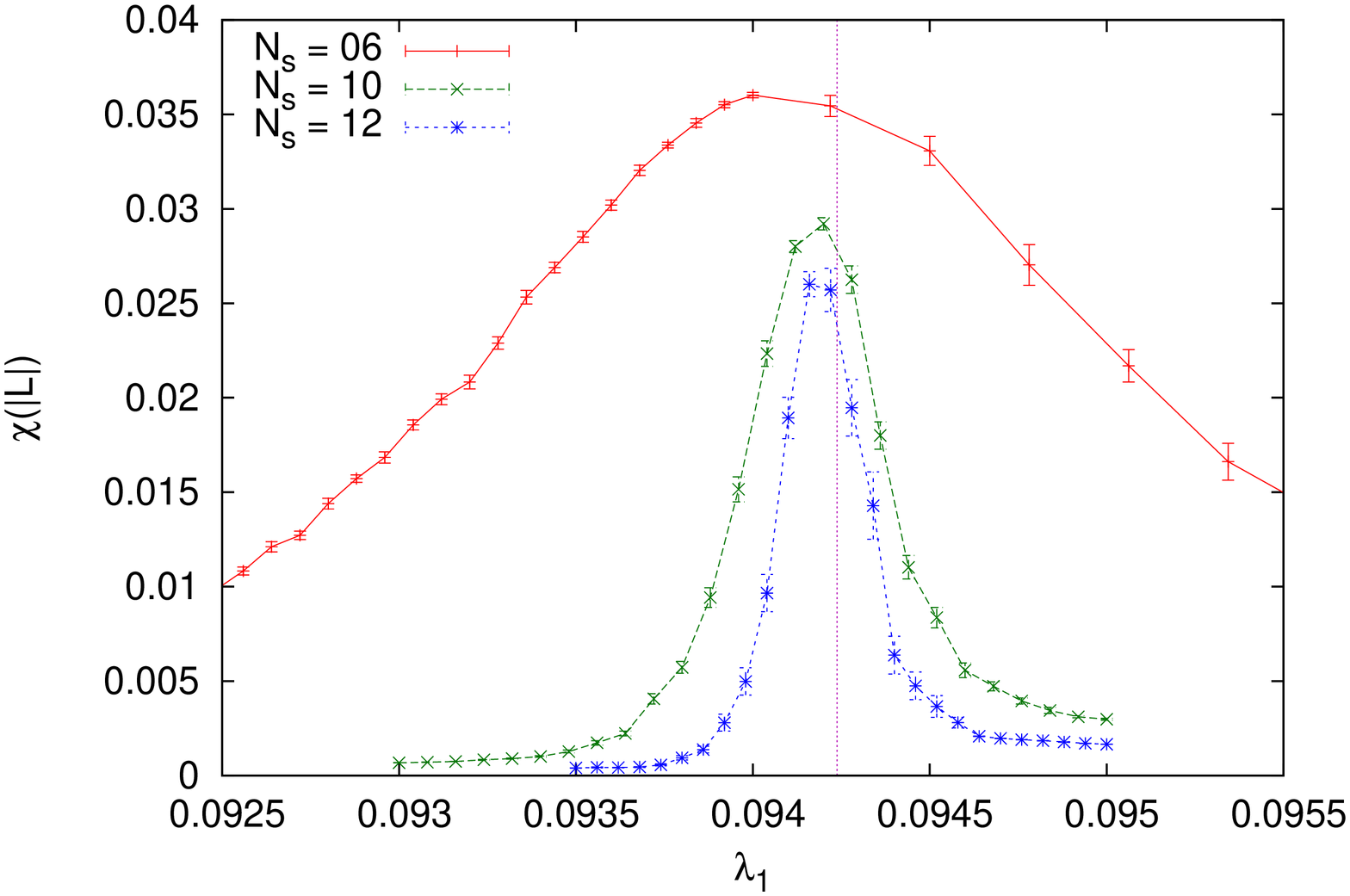}
\caption[]{Left: Distribution of $L$ for small and large $\lambda_1$ on a lattice with $N_s=6$ and $M=1$. Middle, Right: Expectation
value of $|L|$ and its susceptibility. The vertical line marks the infinite-volume transition.}
\label{fig:l}
\end{figure}

Our first task is to establish the phase structure of the effective theory, where we focus 
on the physically interesting case of $SU(3)$. Based on the global 
$Z(3)$ symmetry of the model, one expects spontaneous breaking of that symmetry 
for some critical value of the coupling $\lambda_{1,c}$. Figure \ref{fig:l} shows the behaviour
of the field variable $L$ as a function of $\lambda_1$. As expected from the 4d parent theory,
there is indeed a transition from a disordered or mixed phase, with values of $L$ 
scattering about zero, to an ordered phase at large coupling
where the three $Z(3)$-phases are populated separately.  In the thermodynamic limit, 
one of these vacua will be chosen and the symmetry is broken spontaneously,
$\langle L \rangle=0$ for $\lambda_1<\lambda_{1,c}$ and 
$\langle L \rangle \neq0$ for $\lambda_1>\lambda_{1,c}$. 
Correspondingly, the expectation value of $|L|$ rises abruptly at some critical coupling 
$\lambda_{1,c}$, as shown in figure \ref{fig:l} (middle). On a finite size lattice, the phase
transition is smoothed out,  non-analyticities are approached gradually with growing volume, 
as the figure illustrates.

The general technique to locate the infinite-volume critical coupling, $\lambda_{1,c}$,
is based
on a finite-size scaling analysis. A variety of cubic systems are simulated 
and for each one a pseudo-critical 
point $\lambda_{1,c}(N_s)$ is found.
As a function of lattice side length $N_s$, close enough to the thermodynamic limit
these scale as
\eq
	\lambda_{1,c}(N_s) = \lambda_{1,c} + b N_s^{-1/\nu}\;.
	\label{eq:fss-critical}
\qe
The relevant universal values for the critical index are $\nu = 0.63002$ for 3d Ising
\cite{ising_3d_index_nu}
and $\nu=1/3$ for a first order transition.
In practice we found it sufficient to use spatial volumes $N_s = 6, 
\ldots, 12$ in order to extrapolate to the thermodynamic limit.
In this way, the whole data production can 
be carried out in a handful of days on a modest desktop PC.

The pseudo-critical coupling can be defined in a variety of ways: one
possibility is to employ the energy observable:
\eq
	E = - \frac{1}{\lambda_1} S_\mathrm{eff}'\;,
\qe
where the prime denotes omission of the potential term $V_x$. For the 
$M=1$ model,
this coincides with the usual energy,\footnote{Also in the $\lambda\to 0$ 
limit, at any truncation, this definition
recovers the standard energy term.} while at higher truncations (as well 
as in the non-truncated formulation) 
$S_\mathrm{eff}$ is non-linear in the coupling. For this 
reason, the energy is used only
in the $M=1$ particular case. 
Other natural  observables are based on the
modulus of the Polyakov loop, $|L|$.
We consider the 
Binder cumulant
	\eq
		B(|L|) = 1 - \frac{\avg{|L|^4}}{3\avg{|L|^2}^2} \;,
	\qe
whose minimum defines a pseudo-critical coupling, and the susceptibility
	\eq
		\chi(|L|) = \bigavg{\Big(|L|-\avg{|L|}\Big)^2}\;,
	\qe
whose maximum is taken as pseudo-critical coupling, cf.~figure \ref{fig:l} (right). 
Having various definitions of pseudo-critical
couplings, one can check for the
mutual consistency of the infinite-volume critical points coming from them.

\subsection{Critical coupling and order of the transition for $SU(3)$}

The truncated theories with $M=1,3,5$ were simulated on lattices with spatial sizes 
$N_s=6,8,10$ (plus $N_s=12$ for the $M=1$ theory). 
For each volume, $\sim 30$ values of the couplings are sampled 
by $\sim 10^6$ update sweeps each. Measurements 
were taken every $\sim 30$ updates.

We begin our presentation of results with the $SU(3)$ family of effective theories, where 
we find a first-order transition regardless of the particular
truncation employed. A practical difficulty is the related occurrence of metastabilities 
with extremely long thermalisation times $\propto 
\exp(c N_s^3)$. The local
Metropolis update requires a number of iterations akin to a tunnelling 
time in order to dispel 
configurations with two halves of the system associated with different vacua. 
To take care of this problem,
very long trajectories are inspected by cutting out an initial block of 
configurations until stability is
found in the resulting averages: for instance, a system with size $N_s=16$ 
would require, around criticality,
$\sim 10^6$ update sweeps to thermalise.

\begin{figure}
\begin{center}
	\includegraphics[height=0.5\textwidth, angle=270]{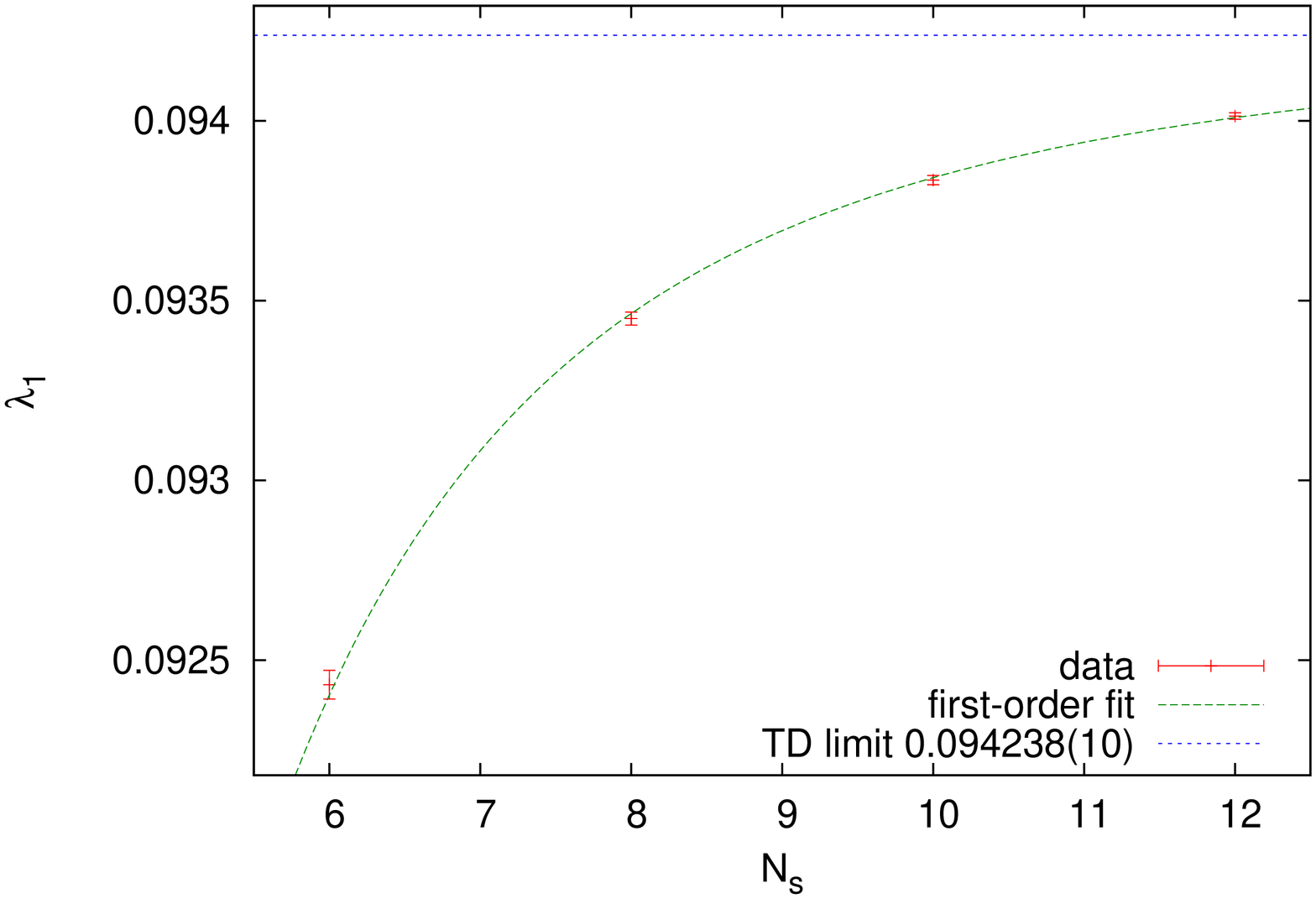}\hspace*{0.2cm}
	\includegraphics[height=0.5\textwidth, angle=270]{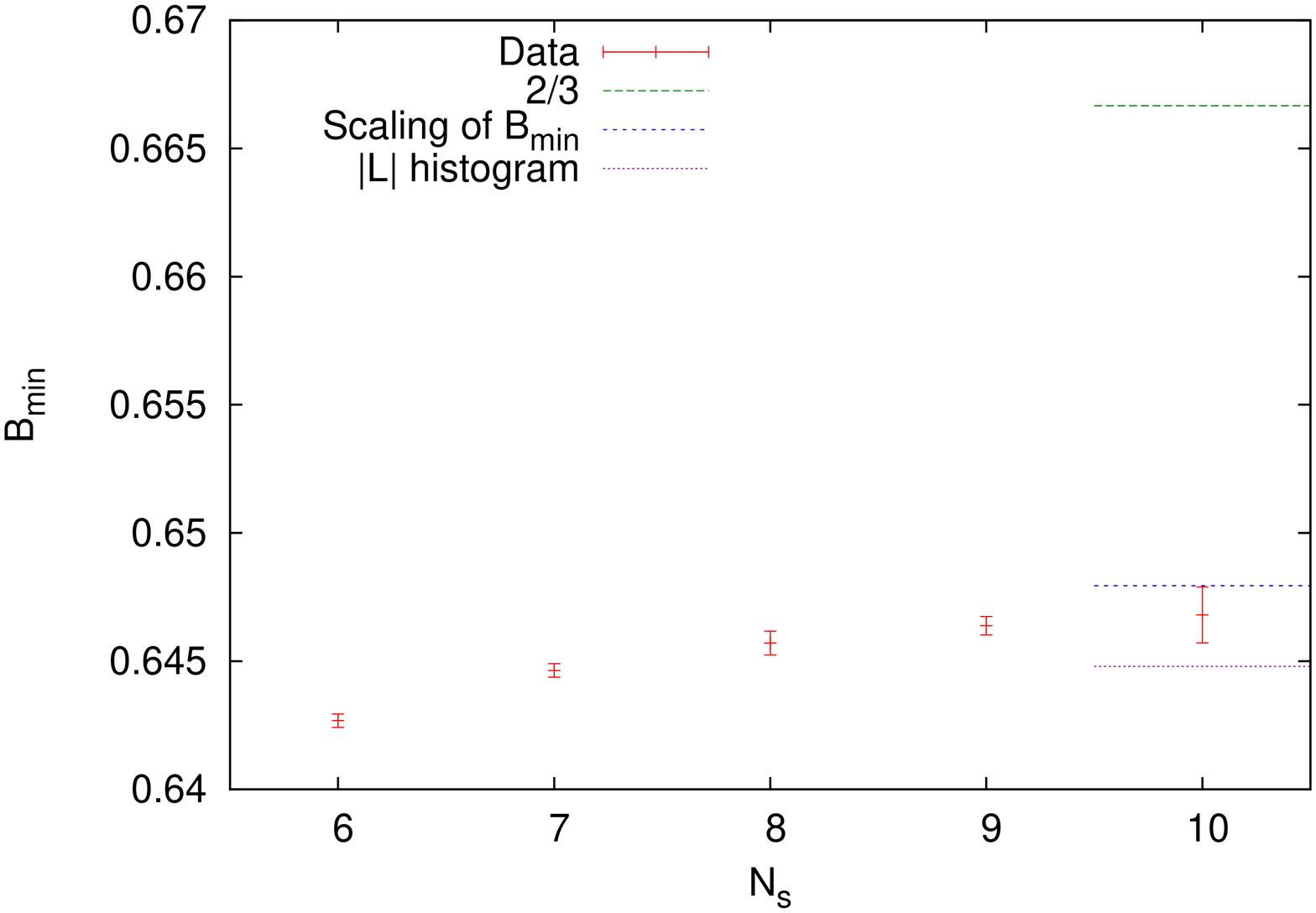}
 	\caption[]{Left: 
 		Position of the minimum of the Binder cumulant $B(E)$ for  $SU(3), M=1$, for different lattice sizes. 
		The horizontal 
		line is the thermodynamic limit resulting from the fit to Eq.~(\ref{eq:fss-critical}).
		Right: Behaviour of $B_\mathrm{min}(N_s)$, along with its thermodynamic limit obtained with the
		$\mathcal{O}(N_s^{-3})$ scaling law and the independent estimate $B_\infty$ from 
		the $|L|$ histogram. Also the second-order limit value $2/3$ is shown.}
	\label{fig:first-order-fss_m1_en}
\end{center}
\end{figure}

\begin{figure}
\begin{center}
	\includegraphics[width=0.5\textwidth, height=0.302\textwidth]{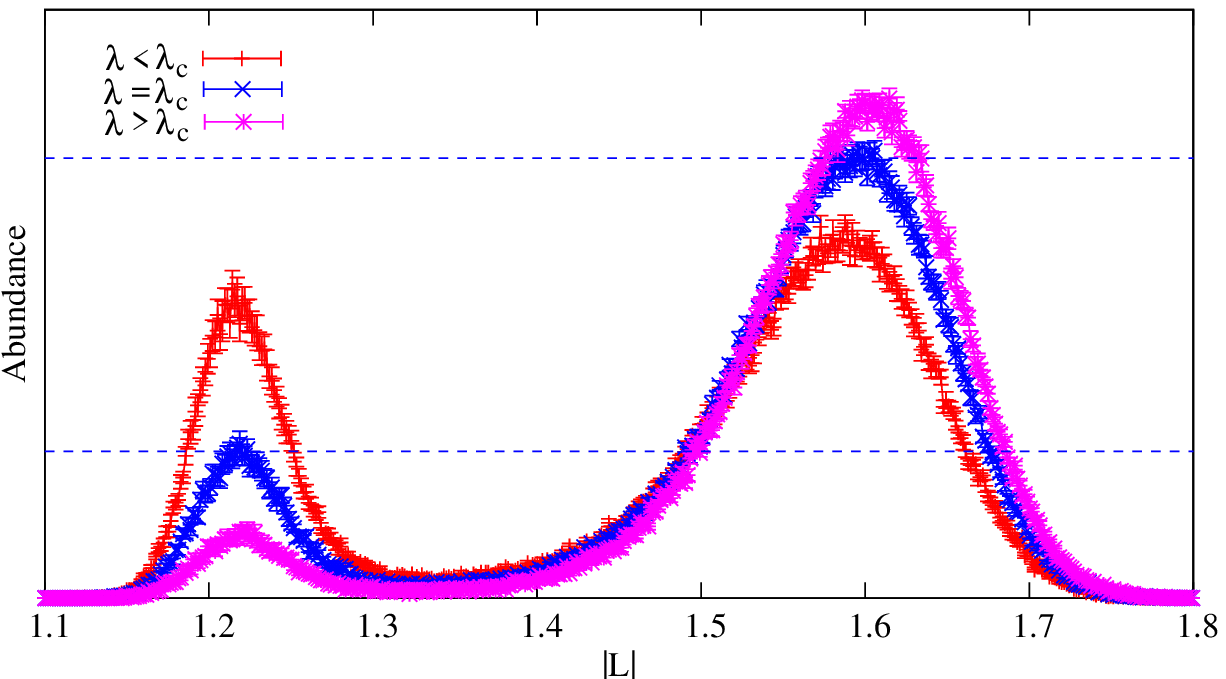}\hspace*{0.2cm}
	\includegraphics[width=0.5\textwidth, height=0.302\textwidth]{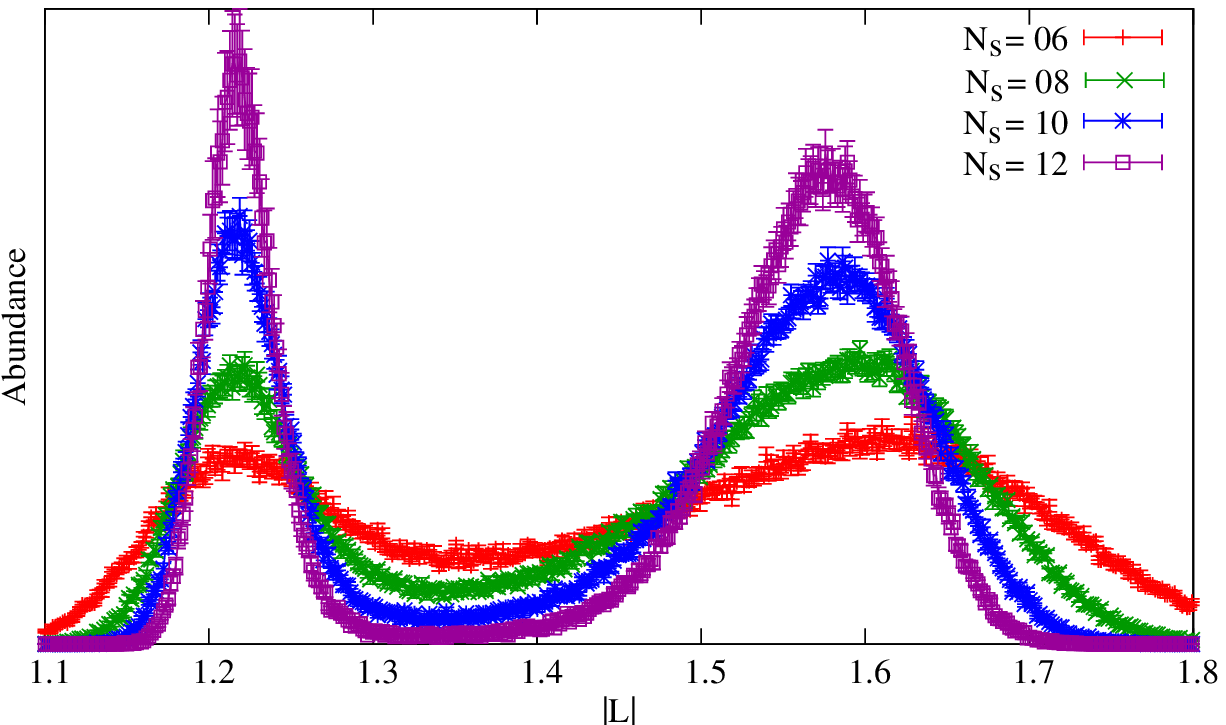}
	\caption[]{Left: Histogram for $|L|$ of the $SU(3)$ theory on $N_s=10$ with $M=1$. The three curves
	have been measured with $\lambda_1$ slightly below, exactly at and slightly above pseudo-criticality.
	The horizontal lines are guides to the eye and mark the peak on the right as being, for the 
	pseudo-critical curve, exactly three times as high as the one on the left.
	Right: Histogram of the $SU(3)$ $|L|$ for four different system sides $N_s$ at the true critical point.
	The size-dependence of the peaks' position was taken into account in estimating
	$B_\infty$ with Eq.~(\ref{bestimate}).}
	\label{fig:double-peak-binder-L}
\end{center}
\end{figure}

First we consider the model with $M=1$.
The first-order nature of the transition is established by fitting the pseudo-critical couplings to the
 scaling law, Eq.~(\ref{eq:fss-critical}), with $\nu=1/3$, as shown in 
figure \ref{fig:first-order-fss_m1_en} (left). The same conclusion is reached when analysing the scaling of the 
minimum $B_\mathrm{min}(N_s)$ of $B(|L|)$, figure \ref{fig:first-order-fss_m1_en} (right): 
this quantity, in the $N_s\to\infty$ limit, approaches the saturation value $2/3$ if and only if 
the transition is second-order, while in case of a first-order point the thermodynamic 
limit is given by \cite{Binder1_lee_kosterlitz}:
\eq
	B_\infty = \frac{2}{3} - \frac{1}{12}\Big( \frac{|L|_1}{|L|_2} - \frac{|L|_2}{|L|_1} \Big)^2 \;,
\label{bestimate}
\qe
where $|L|_1$ and $|L|_2$ are the local maxima of the $|L|$ double-peaked histogram. Inspection of the
$|L|$ distribution shows indeed a two-peaked shape (figure \ref{fig:double-peak-binder-L}) that becomes
narrower and higher at increasing system volumes. It is then possible to give an independent estimate
for $B_\infty$ and compare it with that from the scaling of the $B(|L|)$ minimum.
The comparison (figure \ref{fig:first-order-fss_m1_en}, right) confirms indeed the first-order nature
of the transition, while some slight mismatch (less than two standard deviations) remains
between the two estimates. This is most likely due to the presence of subleading terms in the scaling of
$B_\mathrm{min}$. As demonstrated in \cite{Binder2_billoire_neuhaus_berg}, the scaling law to NLO has the form
\eq
	B_\mathrm{min}(N_s) = B_\infty + B^{(2)} N_s^{-3} + B^{(3)} N_s^{-6}\;,
\qe
and it is necessary to explore very large lattices to observe 
the leading-order scaling.

\begin{figure}
\begin{minipage}{10cm}	
	\includegraphics[height=8.5cm,angle=270]{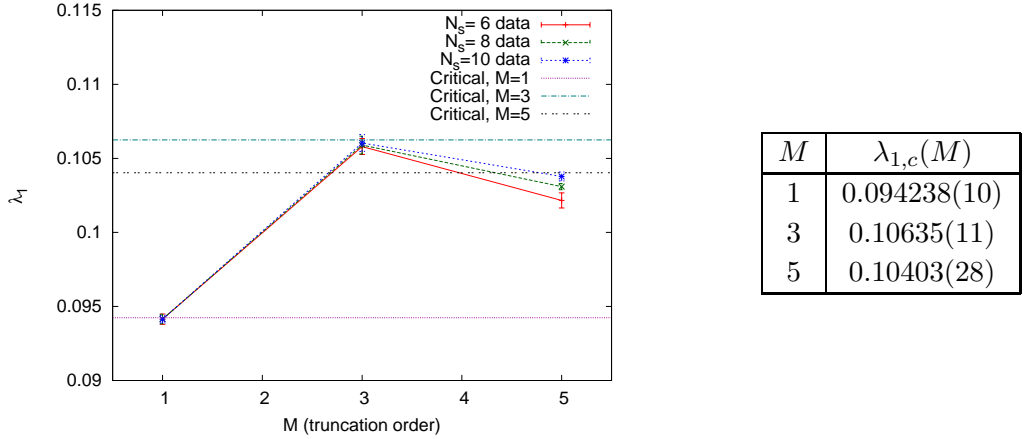}
\end{minipage}	
\begin{minipage}{5cm}
\begin{tabular}{|c|c|}
\hline
$M$ & $\lambda_{1,c}(M)$ \\
\hline
$1$ & $0.094238(10)$ \\
$3$ & $0.10635(11)$ \\
$5$ & $0.10403(28)$ \\
\hline
\end{tabular}
\end{minipage}
 	\caption[]{Critical points in the thermodynamical limit (horizontal lines) and 
	pseudocritical values (data points) on three spatial sizes, for the 
	$SU(3)$ one-coupling model truncated at $M=1,3,5$.}
	\label{fig:su3_m135}
\end{figure}

In the next step we need to investigate the behaviour of the models with higher $M$.
Again we observe first order transitions,
which become 
sharper with increasing $M$. Moreover, finite-size effects are 
stronger for higher $M$, figure \ref{fig:su3_m135}. The critical 
couplings identified for the $M=1,3,5$ effective theories in the thermodynamic limit
are also quoted in figure \ref{fig:su3_m135}.
Judging from these three values, the series seems to be rapidly converging, with only $\sim 3\%$ 
difference between $M=3,5$. The residual difference between this 
estimate and the $M=\infty$ critical coupling is completely subdominant 
compared to the other systematic errors contributing to 
the final results. Also, the direct comparison with the $SU(2)$ case below, 
where the $M=\infty$ data are directly available, supports 
a rapid convergence, figure \ref{fig:su2_m_L8}.

\subsection{Critical coupling and order of the transition for $SU(2)$}

In the case of the $SU(2)$ models, the same analysis was carried out. 
Since here we find 
second-order transitions, the 
metastability problems are absent
and the thermalisation times are orders of magnitude shorter (e.g.~around 
4000 update sweeps at $N_s=16$ around criticality).
This enabled us to study larger system sizes up to $N_s=28$ without 
much computational effort.

\begin{figure}
\begin{center}
	\includegraphics[height=7.2cm,angle=270]{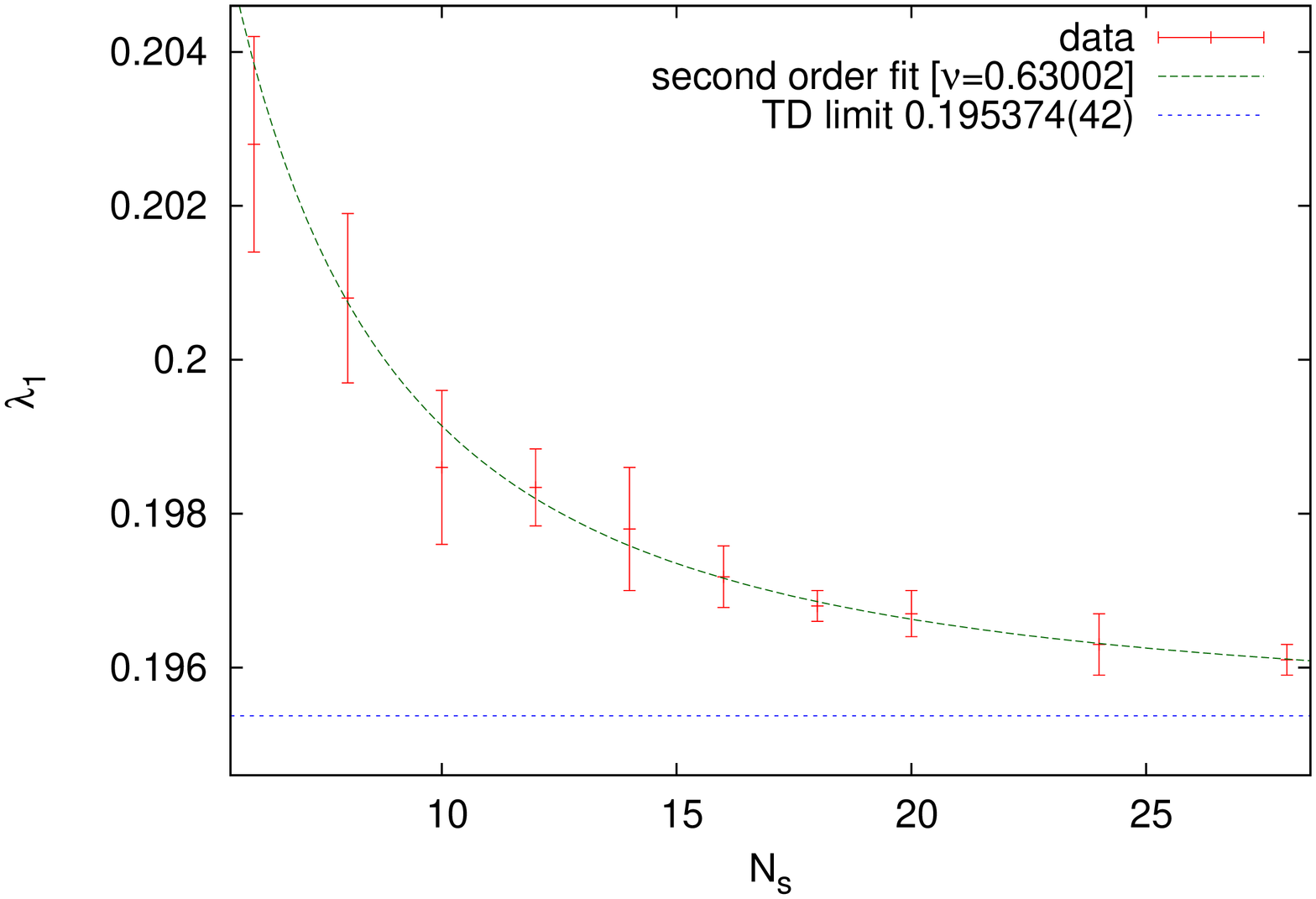}
	\hspace{0.4cm}
	\includegraphics[height=7.2cm, angle=270]{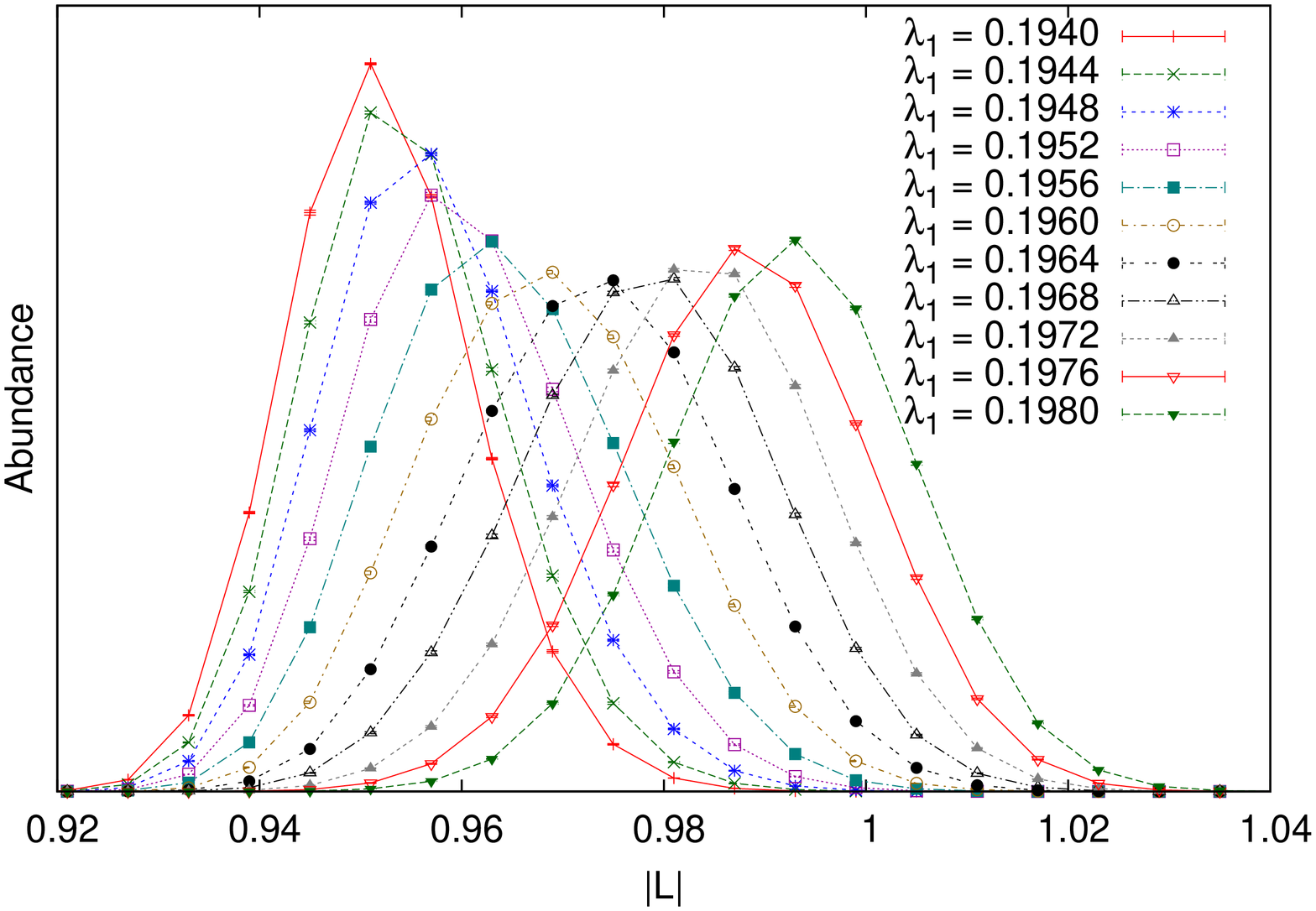}
 	\caption[]{$SU(2)$  
	with $M=1$; Left: Location of maximum of 
	the $E$-susceptibility, fitted to the 3D Ising scaling law. Right: Histogram for $|L|$ at $N_s=20$.}
	\label{fig:su2-second-order-plots}
\end{center}
\end{figure}

\begin{figure}
\begin{center}
	\includegraphics[height=7.2cm,angle=-90]{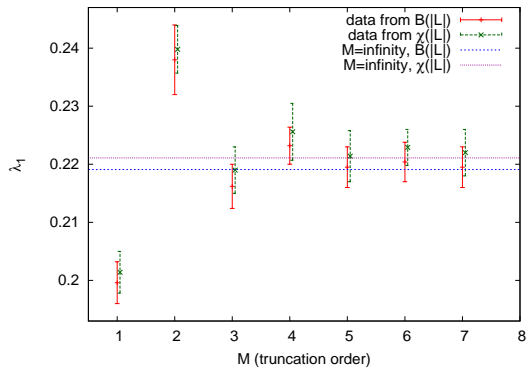}
 	\caption{$\lambda_{1,c}(N_s=8)$ for $SU(2)$ and
	 various truncations. Estimates 
	from $B(|L|), \chi(|L|)$ are shown; 
	horizontal lines mark the values
	for the model without truncations $(M=\infty)$.}
	\label{fig:su2_m_L8}
\end{center}
\end{figure}

The same approaches as before clearly exhibit the second order nature of the 
transition,  using the scaling law for
the pseudo-critical points and the histogram for $|L|$ 
(see figure \ref{fig:su2-second-order-plots}), as well as the corresponding Binder 
cumulant
thermodynamic limit which in this case reaches smoothly the value $2/3$.
Figure \ref{fig:su2-second-order-plots} shows the results for the $M=1$ model.
However, as indicated before, in the case of $SU(2)$ we can directly work with the full
$M=\infty$ model and thus check for the systematic errors when a finite $M$ truncation is
used. This is shown in figure \ref{fig:su2_m_L8}, explicitly we find
\eq
	\lambda_{1,c}(M=1)=0.195374(42)\;\;;\;\;\lambda_{1,c}(M=\infty)=0.21423(70).
\qe
Indeed we observe rapid convergence
on the $M=\infty$ results, in accord with our findings for $SU(3)$. 
We thus conclude that 
the truncation in $M$ does not affect the results significantly and proves to be a viable way of 
dealing with the numerical difficulties
discussed above.

\subsection{Two-coupling models for $SU(3)$}

In this section we study the influence of including a second coupling. We start with the leading next-to-nearest-neighbour interaction,
specified by the model with a second coupling $\lambda_2$, Eq.~(\ref{eq:2coupling_su2}).
For $SU(3)$, the two-coupling 
partition function reads
\eq
	Z = \Big(\prod_x \int \de L_x\Big) \prod_{<ij>}(1+2\lambda_1 \Real L_i L^*_j) \prod_{[kl]}(1+2\lambda_2 \Real L_k L^*_l)
	e^{\sum_x V_x}\;.
	\label{eq:2coupling_su3}
\qe
As discussed previously, we neglect a third coupling related to next-to-nearest-neighbour 
interaction between loops at distance $2a$ since it starts 
at higher order than $\lambda_2$. 

In a natural extension of the critical point of
the one-coupling model, 
there is now a
critical line $\lambda_{1,c}(\lambda_2)$ separating the ordered from the disordered phase 
in the two-dimensional parameter space. However, not all points on this line
are related to the physics of the 4d thermal theory.
Once a particular $N_\tau$ is fixed,
both $\lambda_i$ depend on the expansion parameter $u(\beta)$ alone. Eliminating $u$, 
the curves $\lambda_1^{(N_\tau)}(\lambda_2)$ can be constructed, which represent
the parameter space describing the physics of the 4d thermal model for a given $N_\tau$.
The critical couplings relevant for us 
are thus the intersections between the critical line of the effective two-coupling model and the curves 
specifying the map to a particular $N_\tau$ lattice.

We remark that the second factor in the partition function, 
Eq.~(\ref{eq:2coupling_su3}), has the same 
kind of ``sign problem'' as the first one, calling 
for an analogous 
solution. Some attention needs to be paid to truncating the 
expansion of the two logarithms in a consistent way. 
We include all terms which contribute to the desired power of $u$ from both 
expansions, for all values of $N_\tau$. Based on our previous results, 
we want to keep terms up to $M=3$ for the $\lambda_1$ part, and thus consider the leading
 second coupling term $M=1$, which we denote with $(M_1,M_2)=(3,1)$. Corrections by higher powers
 $M_2$ are formally of higher order in $u$, and we have checked numerically that their effect is completely negligible. 

\begin{figure}
\begin{center}
	\includegraphics[height=7.2cm,angle=270]{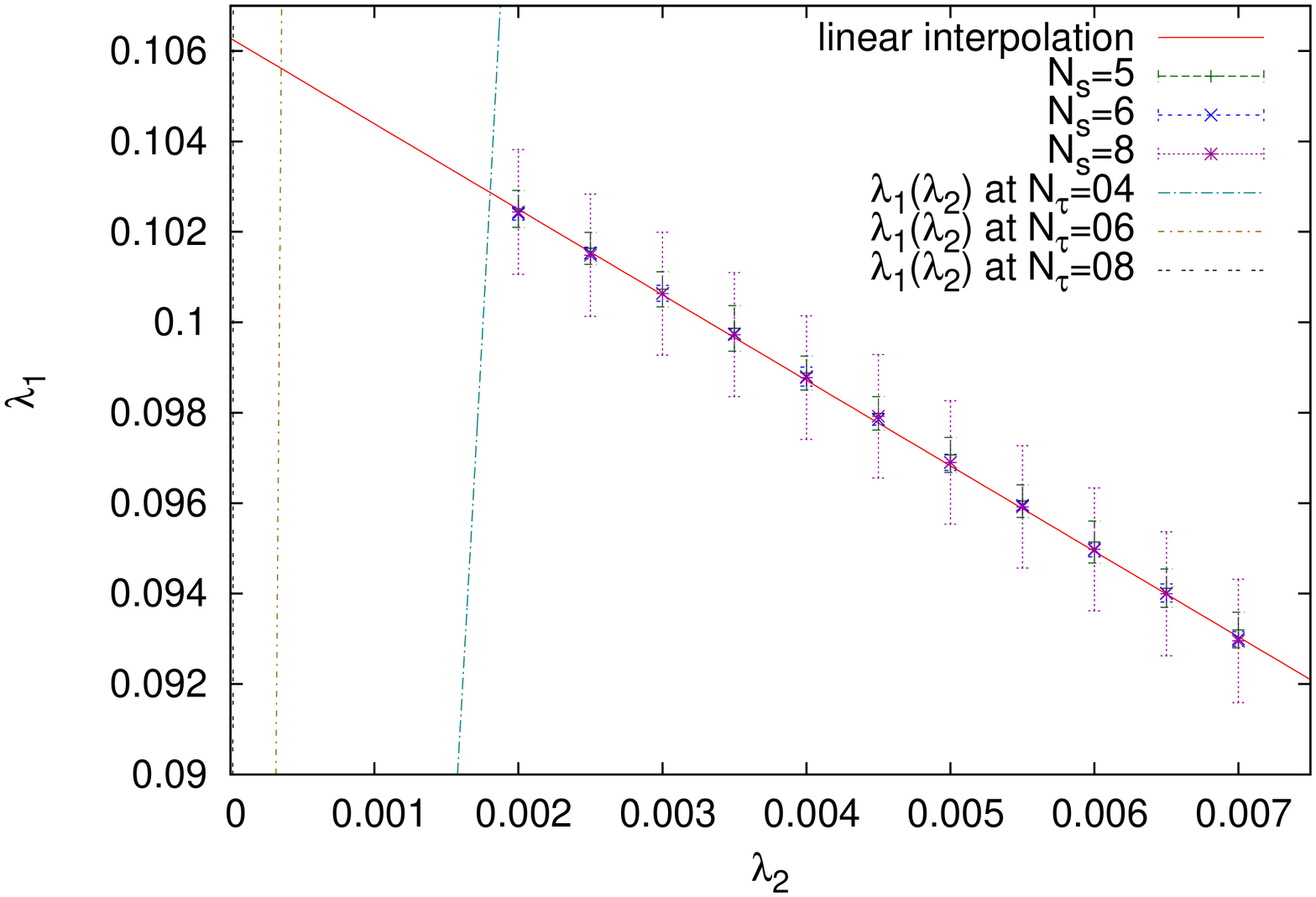}
	\includegraphics[height=7.2cm,angle=270]{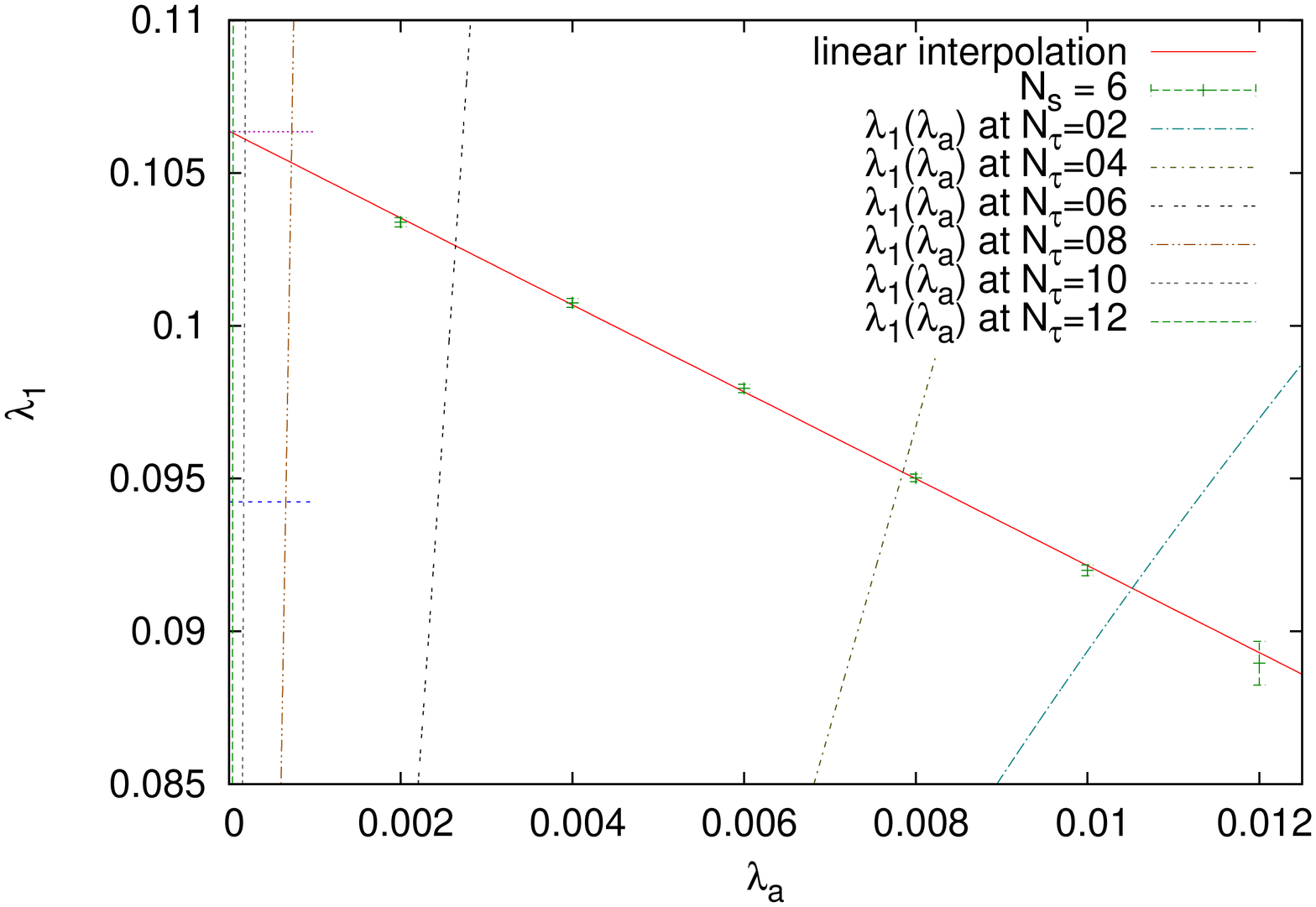}
 	\caption[]{Critical line in the $SU(3)$ two-coupling space, determined from $\chi(|L|)$.
	Dashed lines give the parameter space representing a 4d theory with fixed $N_\tau$.
	Left: $(\lambda_1,\lambda_2)$. Right: $(\lambda_1,\lambda_a)$.}
	\label{fig:critline-2coupling}
\end{center}
\end{figure}

In order to locate the critical line $\lambda_{1,c}(\lambda_2)$, we fix  11 different 
values of $\lambda_2$ in the range $[0.002,0.007]$ and scan in
$\lambda_1$ in order to identify  
the pseudo-critical points, 
which can then be extrapolated to the thermodynamic limit.
It turns out that, on a given volume, $\lambda_{1,c}(\lambda_2)$ is well described by a linear 
interpolation, cf.~figure \ref{fig:critline-2coupling} (left). 
Moreover, the inclusion of the second coupling appears to
diminish finite size effects. Since all three volumes shown in figure \ref{fig:critline-2coupling} (left)
are consistent within errors, 
we quote the result from the largest lattice for the critical line,
\eq
	\lambda_{1,c} = a+ b \lambda_2 \quad \mbox{with}\quad
	a = \phantom{+}0.10628(8),\;
	b = -1.891(4)\;.
\qe
The value of $a$ is consistent with the $M=3$ critical point in the one-coupling theory.

Also shown in figure \ref{fig:critline-2coupling} (left) are the lines $\lambda_1^{(N_\tau)}(\lambda_2)$
for $N_\tau=4,6,8$. 
One observes that they rapidly accumulate towards $\lambda_2 \simeq 0$. 
Only for the lowest 
values of $N_\tau$, corresponding to coarse lattices in the 4d theory, does the adoption of the two-coupling model make any 
difference. For finer lattices, of interest for continuum physics, the results are within
statistical errors indistinguishable from the the simpler one-coupling theory. 

Next, we are considering a two coupling theory with $\lambda_1,\lambda_a$.
In this case the partition function reads
\eq
        Z = \Big(\prod_x \int \de L_x\Big) \prod_{<ij>}(1+2\lambda_1 \Real L_i L^*_j)
  \prod_{<ij>}e^{\lambda_a (\tr^{(a)}W_i)(\tr^{(a)}W_j)}
        e^{\sum_x V_x}\;.
\qe
Here $\tr^{(a)}$ denotes the trace in the adjoint representation, $\tr^{(a)} W = |\tr W|^2-1$, which 
can be used to rewrite the action in terms of fundamental loops. For the numerical evaluation, we
again expand the $\lambda_1, \lambda_a$ terms to $(M_1=3,M_2=1)$, and proceed in complete
analogy as in the case discussed above. The result for the critical line in this two-coupling space is
 shown in figure \ref{fig:critline-2coupling} (right). Here we find
 \eq
	\lambda_{1,c} = a+ b \lambda_a \quad \mbox{with}\quad
	a = \phantom{+}0.10637(15),\;
	b = -1.422(22)\;.
\qe
Once more, the set of curves intersecting the critical line correspond to lines of fixed $N_\tau$ in
the 4d theory. We observe that $\lambda_a$ has slightly larger effect than $\lambda_2$ at fixed $N_\tau$,
in accord with the fact that it starts at lower order in $u$, cf.~Eq.~(\ref{adjlam}). Nevertheless, its
influence is smaller than that of the strong coupling truncation in $\lambda_1$, as we shall see, and 
hence negligible at this order.

\section{Mapping back to 4d Yang-Mills}
\label{sec:results}

Having established the critical couplings for our effective theories and tested their 
reliability, we are now ready to map them back to the original thermal Yang-Mills theories
by using Eqs.~(\ref{eq_lambda}, \ref{eq_lambda2}).
In Tables \ref{tab:su2_betas}, \ref{tab:su3_betas} we collect the values for 
the critical gauge couplings, $\beta_c$, obtained in this way from the effective
theories and compare them to the values obtained from
simulations of the full 4d theories for $SU(2), SU(3)$, respectively. 

\begin{figure}
\begin{center}
	\includegraphics[height=7.2cm,angle=270]{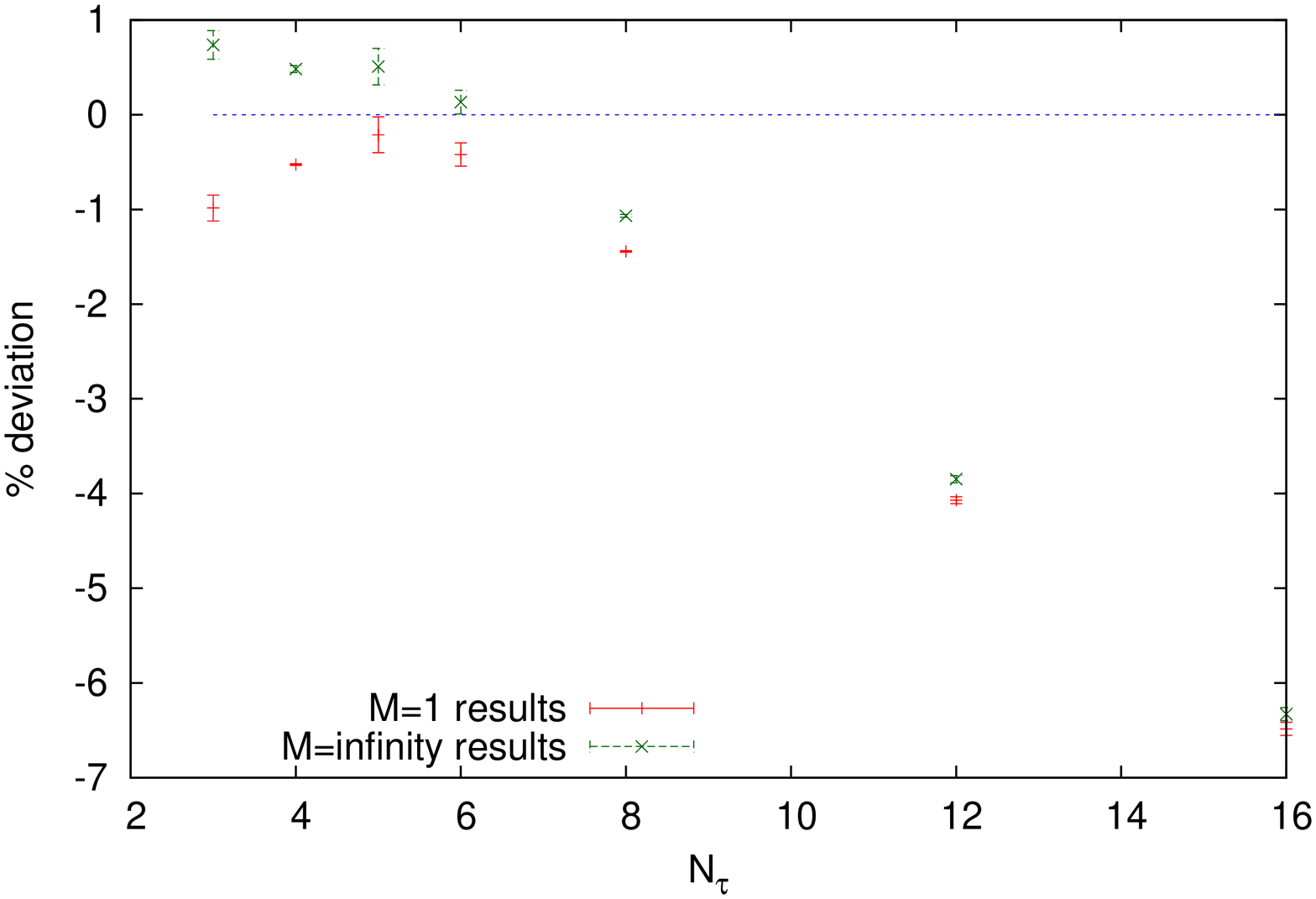}	
	\hspace{0.4cm}
 		\includegraphics[height=7.2cm,angle=270]{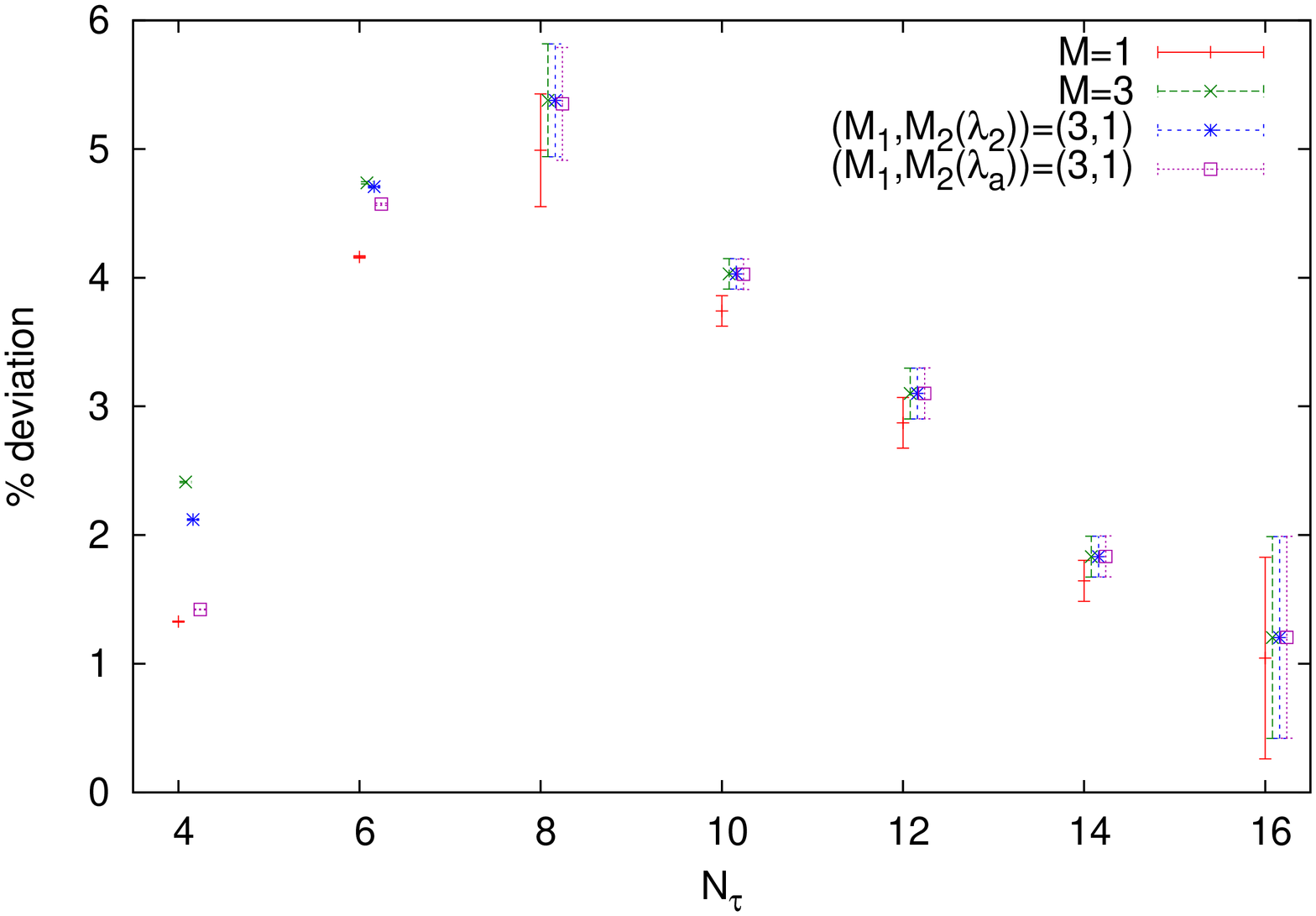}
	\caption[]{Relative error of $\beta_c$ predicted by the effective theories when compared 
	to simulations of the 4d theories, for $SU(2)$ (left) and $SU(3)$ (right).}
	\label{fig:betas}
\end{center}
\end{figure}

\begin{table}[t]
\begin{center}
\begin{tabular}{|c||c|c||c|}
\hline
	$N_\tau$ & $M=1$ & $M=\infty$ & $\mbox{4d YM}$ \\
\hline
	3    &    2.15537(89)   &     2.1929(13)   &     2.1768(30) \\
	4    &    2.28700(55)   &     2.3102(08)   &     2.2991(02) \\
	5    &    2.36758(40)   &     2.3847(06)   &     2.3726(45) \\
	6    &    2.41629(32)   &     2.4297(05)   &     2.4265(30) \\
	8    &    2.47419(22)   &     2.4836(03)   &     2.5104(02) \\
	12   &    2.52821(14)   &     2.5341(02)   &     2.6355(10) \\
	16   &    2.55390(10)   &     2.5582(02)   &     2.7310(20) \\
\hline
\end{tabular}
\caption{Critical couplings $\beta_c$  for $SU(2)$ from two effective 
theories compared to simulations of the 4d theory \cite{fingberg_heller_karsch_1993,bo,Ve}).}
\label{tab:su2_betas}
\end{center}
\end{table}~\begin{table}[t]
\begin{center}
%\hspace*{-1cm}
\begin{tabular}{|c||c|c|c|c||c|}
\hline
	$N_\tau$ & $M=1$ & $M=3$ & $M_1,M_2(\lambda_2)=3,1$ &
	$M_1,M_2(\lambda_a)=3,1$ & $\mbox{4d YM}$\\
\hline
	4	&  5.768 & 5.830 & 5.813 & 5.773 &5.6925(002) \\
	6	&  6.139 & 6.173 & 6.172 & 6.164 &5.8941(005) \\
	8	&  6.300 & 6.324 & 6.324 & 6.322 &6.0010(250) \\
	10	&  6.390 & 6.408 & 6.408 & 6.408 &6.1600(070) \\
	12	&  6.448 & 6.462 & 6.462 & 6.462 &6.2680(120) \\
	14	&  6.488 & 6.500 & 6.500 & 6.500 &6.3830(100) \\
	16	&  6.517 & 6.528 & 6.528 & 6.528 &6.4500(500) \\
\hline
\end{tabular}
\caption{Critical couplings $\beta_c$  for $SU(3)$ from different effective 
theories compared to simulations of the 4d theory 
\cite{Kogut_et_al_1983,fingberg_heller_karsch_1993}).}
\label{tab:su3_betas}
\end{center}
\end{table}
The agreement is remarkable in all cases, with the relative error of the effective theory 
results compared to the full ones shown in figure \ref{fig:betas}. 
The comparison of alternative truncations of the logarithm shows once more that it has almost no 
influence on the accuracy of the final result, as described earlier.
Interestingly, there appears to be
a `region of best agreement', with the deviation growing both for small and large $N_\tau$.
We ascribe this to the fact that there are two competing systematic errors, as discussed earlier:
the validity of the strong coupling series for a given coupling $\lambda_i$ is better
the smaller $\beta$ and hence $N_\tau$, whereas the truncation of the next-to-nearest neighbour
interactions gains validity with growing $N_\tau$. In particular in the case of $SU(3)$, there appears
to be a cancellation of the two kinds of systematics, rendering the effective description better for the original theory on finer lattices.

\begin{figure}
\begin{center}
	\includegraphics[width=5.0cm,angle=270]{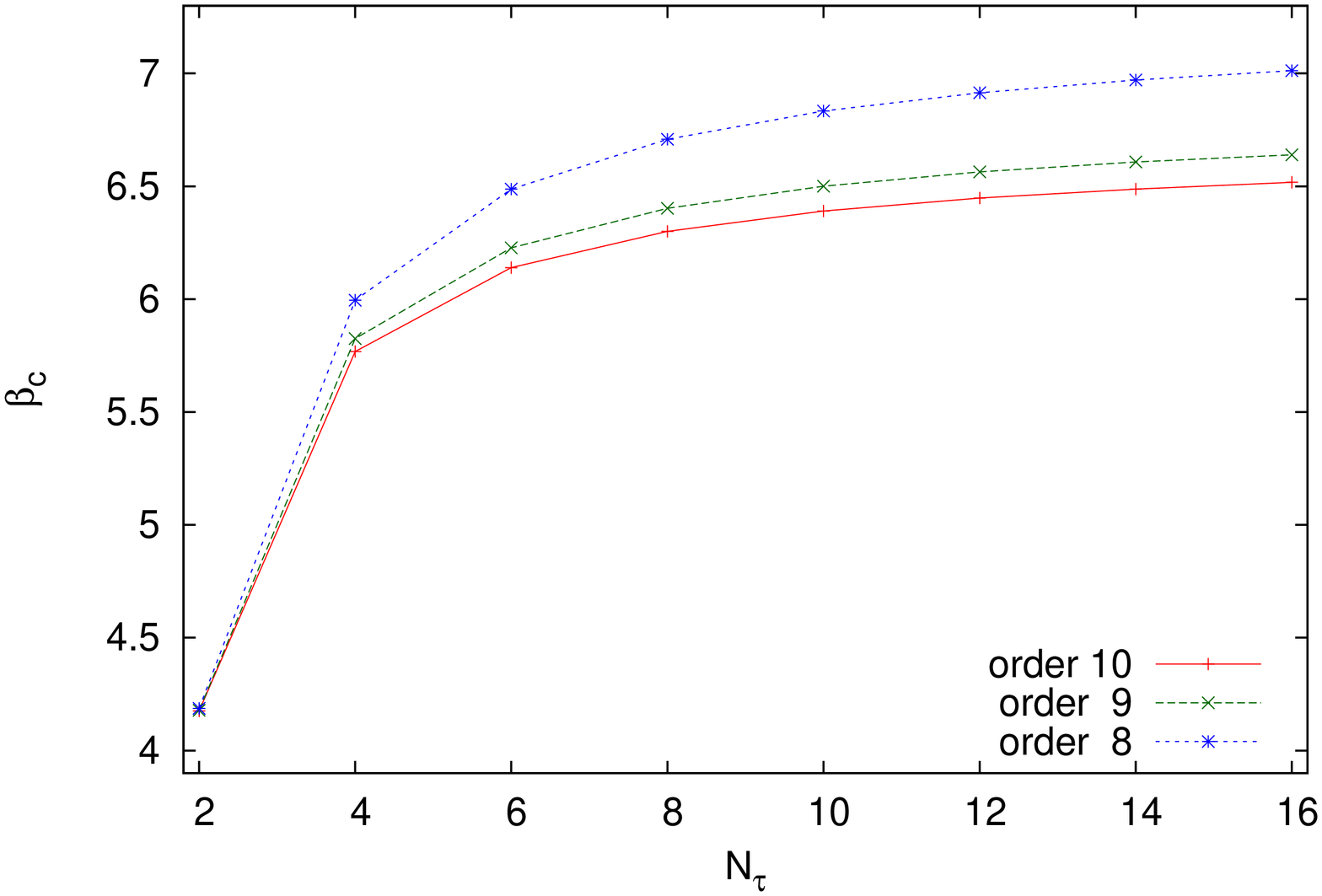}
	\includegraphics[height=7.0cm, angle=270]{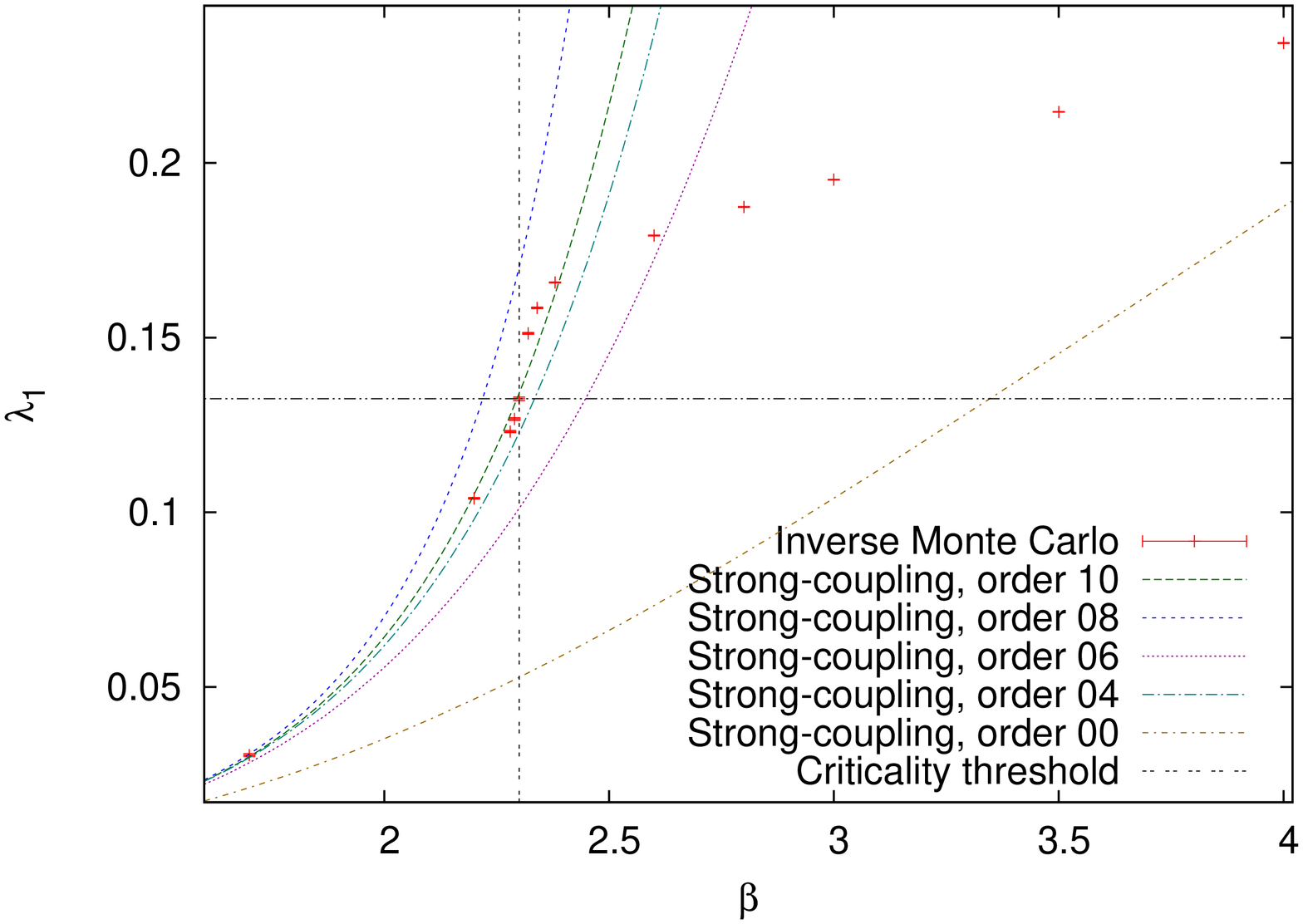}
 	\caption[]{Left: $\beta_c$ for $SU(3)$ from $\lambda_{1,c}(\beta)$  evaluated to
	8th, 9th and 10th order, Eqs.~(\ref{eq_lambda2}).
	Right: $\lambda_1(\beta)$ for $SU(2)$ and $N_\tau=4$ from strong-coupling 
	expansion and inverse Monte 
	Carlo \cite{Heinzl:2005xv}.}
	\label{fig:su3_systematics}
\end{center}
\end{figure}

In order to asses the systematics of the strong coupling series, figure \ref{fig:su3_systematics} (left) 
shows the difference in $\beta_c$ based on the one-coupling model when series of different depth
are used for $\lambda_1(\beta)$. Satisfactory convergence behaviour is observed. 
For the case of $SU(2)$ a comparison with a non-perturbative derivation of the effective theory
is possible. Figure \ref{fig:su3_systematics} (right) compares $\lambda_{1,c}$ 
in terms of our strong coupling series with a determination by inverse Monte Carlo 
\cite{Heinzl:2005xv} (see \cite{Wozar:2007tz} for $SU(3)$). 
Again, we observe good convergence of the strong coupling series to the full result. However,
these plots also illustrate that the error due to truncation of the strong coupling series of the
first coupling is larger than the neglect of higher couplings.
We also attempted to improve the convergence of the series with Pad\'e analysis, which however 
was not particularly successful, apart from confirming the order of magnitude of the estimated
systematic uncertainties.

Figure \ref{fig:su3_systematics} (right) illustrates the range of validity of the strong coupling
derivation of the effective couplings. The non-perturbatively determined 
$\lambda_1(\beta)$ appears to change curvature at $\beta_c$, whereas the estimates based
on the strong coupling series do not. This is consistent with $\beta_c$ marking the radius of 
convergence also for the series expansion of the effective coupling $\lambda_1$. Thus, the
inverse Monte Carlo approach has a wider range of validity whereas the series approach
furnishes analytically known mappings between the full and effective theories.

\section{Conclusions}
\label{sec:conclusions}

Employing strong coupling expansions, we have derived a dimensionally reduced, centre-symmetric 
effective description for lattice pure gauge theories at finite temperature. The effective theory 
is formulated in terms of scalar Polyakov loop variables and does not involve any 
matrix degrees of freedom. 
Contrary to earlier derivations based on the lattice theory,  
we have included spatial plaquettes  systematically. 
The effective theory has an infinite number of interaction terms with increasing degree
of non-locality. Each of its couplings
can be computed as a series in the original lattice gauge coupling $\beta$ for any value of $N_\tau$.
We have explicitly calculated up to seven terms in the nearest-neighbour interaction and up to
three terms in the next-to-nearest neighbour interactions. Furthermore, we have shown
that other interaction terms are formally of higher order than the calculated ones.
These calculations can be systematically improved in the future, if desired.

In the second part of the paper, we have performed detailed numerical studies of the 
effective theory for the cases of $SU(2)$ and $SU(3)$, focusing on the description of its 
order/disorder phase transition, which is related to the 
deconfinement transition of the 4d thermal theories. 
We computed the critical couplings and found the transition 
to correspond to 3d Ising for $SU(2)$ and first order for
$SU(3)$. Explicit numerical checks with different approximations show that 
next-to-nearest neighbour couplings have negligible numerical effects, and furthermore
demonstrate good convergence behaviour of the strong coupling series.
Using the analytic expressions calculated earlier, the critical couplings can
be converted to values for $\beta_c(N_\tau)$ predicting the deconfinement transition in the 
original 4d thermal theories. For lattices with $N_\tau=4-16$, we found no more than 6\% deviations from
those calculated in simulations of the 4d theories.
  
In conclusion, we have given a successful description of the deconfinement transition
of 4d $SU(N)$ Yang-Mills theories in terms of a dimensionally reduced $Z(N)$-model derived
by a strong coupling expansions. Future work might push for increased precision by inclusion of higher orders, which would entail more interaction terms and higher representation loops. Our model
might also be of interest for numerical investigations of the deconfinement transition for $SU(N)$ with
$N>3$, cf.~\cite{largen} and references therein.
However, we believe at this stage it would be most interesting to extend this approach 
towards physical QCD by including fermions and finite baryon density, e.g.~by means of a
hopping parameter expansion \cite{Langelage:2009jb,Langelage:2010yn}. 

% \paragraph{Acknowledgements}
\acknowledgments
We thank Rob Pisarski for discussions and valuable comments on the manuscript.
S.~L. and O.~P. are partially supported by the German BMBF grant \textit{FAIR theory: the QCD
phase diagram at vanishing and finite baryon density},  06MS9150, and by
the Helmholtz International Center for FAIR within the LOEWE program of the State of Hesse. 
J.~L.~acknowledges financial support by the EU project \textit{Study of Strongly interacting 
Matter}, No. 227431, and by the BMBF under the project \textit{Heavy Quarks as
 a Bridge between Heavy Ion Collisions and QCD}, 06BI9002.

\newpage

\begin{titlepage}
\setcounter{page}{20}

\begin{flushright}
%BI-TP-2010/32
\end{flushright}
\begin{centering}
\vfill

{\bf\Large Erratum: Centre symmetric 3d effective actions for thermal SU(N) 
Yang-Mills from strong coupling series}

\vspace{0.8cm}
 
Jens Langelage$^1$, Stefano Lottini$^2$ and Owe Philipsen$^2$

\vspace{0.3cm}
{\em 
$^1$ Fakult\"at f\"ur Physik, Universit\"at Bielefeld, \\
33501 Bielefeld, Germany}

\vspace{0.3cm}
{\em $^2$
Institut f\"ur Theoretische Physik, Goethe-Universit\"at Frankfurt,\\
Max-von-Laue-Str. 1, 60438 Frankfurt am Main, Germany}
\vspace*{0.7cm}
 
%\begin{abstract}
%We derive three-dimensional, $Z(N)$-symmetric effective actions in terms of Polyakov loops 
%by means of strong coupling expansions, starting from thermal $SU(N)$ Yang-Mills theory in four
%dimensions on the lattice. An earlier action in the literature, corresponding to the (spatial) strong coupling limit, 
%is thus extended by several higher orders, as
%well as by additional interaction terms.
%We provide analytic mappings between the couplings of the effective theory
%and the parameters $N_\tau,\beta$ of the original thermal lattice theory, which can be systematically
%improved. We then investigate the deconfinement transition for the cases 
%$SU(2)$ and $SU(3)$ by means of Monte Carlo simulations of the effective theory.
%Our effective models correctly reproduce second order 3d Ising and first order phase transitions, respectively.
%Furthermore, we calculate the critical couplings $\beta_c(N_\tau)$ and find agreement with results from  
%simulations of the 4d theory at the few percent level for $N_\tau=4-16$. 
%\end{abstract}
\end{centering}

\noindent
\vfill
\noindent

\end{titlepage}

\setcounter{page}{21}
\section*{Corrected integral measure for $SU(3)$}
\label{sec:numerical_ERRATA}

The $SU(3)$ effective theories derived by strong coupling expansions in
\cite{Langelage:2010yr} have a path integral representation
 with a measure given in terms of traced Polyakov loops, 
 $L_i=L(\vec{x}_i)$,
 \eq
	Z = \Big(\prod_x \int \de L_x\Big) e^{-S_\mathrm{eff}}
		\;,\;
	S_\mathrm{eff} = -\sum_{<ij>} \log(1+2\lambda_1 \Real L_i L^*_j)-
\sum_x V_x\;.
\qe
The potential term in $S_\mathrm{eff}$ is the Jacobian induced by the Haar measure of 
the original group integration.
For our parametrisation $L(\theta,\phi)$, eq.~(3.2) in \cite{Langelage:2010yr}, 
eq.~(3.4) of \cite{Langelage:2010yr} is incorrect. The corrected formula reads
\eq
	\int \de L_x \;e^{V_x} = \int_{-\pi}^{+\pi}\de\theta_x\int_{-\pi}^{+\pi}\de\phi_x \; e^{2V_x}\;,
\qe
with $V_x$ given in eq.~(3.3); see also eq.~(18) in \cite{Wozar:2007tz}.

The  corresponding corrected effective theories investigated numerically are
those with (1) one-coupling, (1,2) nearest- and next-to-nearest-neighbour couplings,
(1,a) funda\-men\-tal- and adjoint-nearest-neighbours (summations on $<,>$ are on nearest-neighbours, 
those on $[,]$ are on
next-to-nearest neighbours):
\eqa
	Z_{(1)} &=& \underbrace{\prod_x \int \de \theta_x \int \de \phi_x}_{\int \mathcal{D}[\theta,\phi]}
			\underbrace{\prod_x \Big(27-18|L_x|^2+8\Real(L_x^3)-|L_x|^4\Big)}_{\prod_x \exp(2V_x)}
		\prod_{<i,j>}(1+2\lambda_1 \Real L_i L^*_j) \nonumber \\
	Z_{(1,2)} &=&\int \mathcal{D}[\theta,\phi]
		\prod_x \exp(2V_x)
		\prod_{<i,j>}(1+2\lambda_1 \Real L_i L^*_j)
		\prod_{[k,l]}(1+2\lambda_2 \Real L_k L^*_l) \nonumber \\
	Z_{(1,a)} &=&\int \mathcal{D}[\theta,\phi] 
		\prod_x \exp(2V_x)
		\prod_{<i,j>}(1+2\lambda_1 \Real L_i L^*_j)
		\prod_{<m,n>}[1+\lambda_a (|L_m|^2-1) (|L_n|^2-1)] \nonumber \;\;.
\qea
Clearly, this change in the effective action affects all numerical results obtained in 
\cite{Langelage:2010yr} for the case of $SU(3)$. It turns out that the correction
actually facilitates the simulations while modifying the results at the few percent level, such
that the conclusions given in \cite{Langelage:2010yr} remain the same. However, for completeness 
we provide corrected plots obtained from the actions above in the following section (see also \cite{EQCD_Proceeding_2011}). 
All formulae and numerical results for the case of
$SU(2)$ in \cite{Langelage:2010yr} are correct.

\section*{Corrected numerical results}

\begin{figure}
\begin{center}
\includegraphics[height=5.5cm,angle=270]{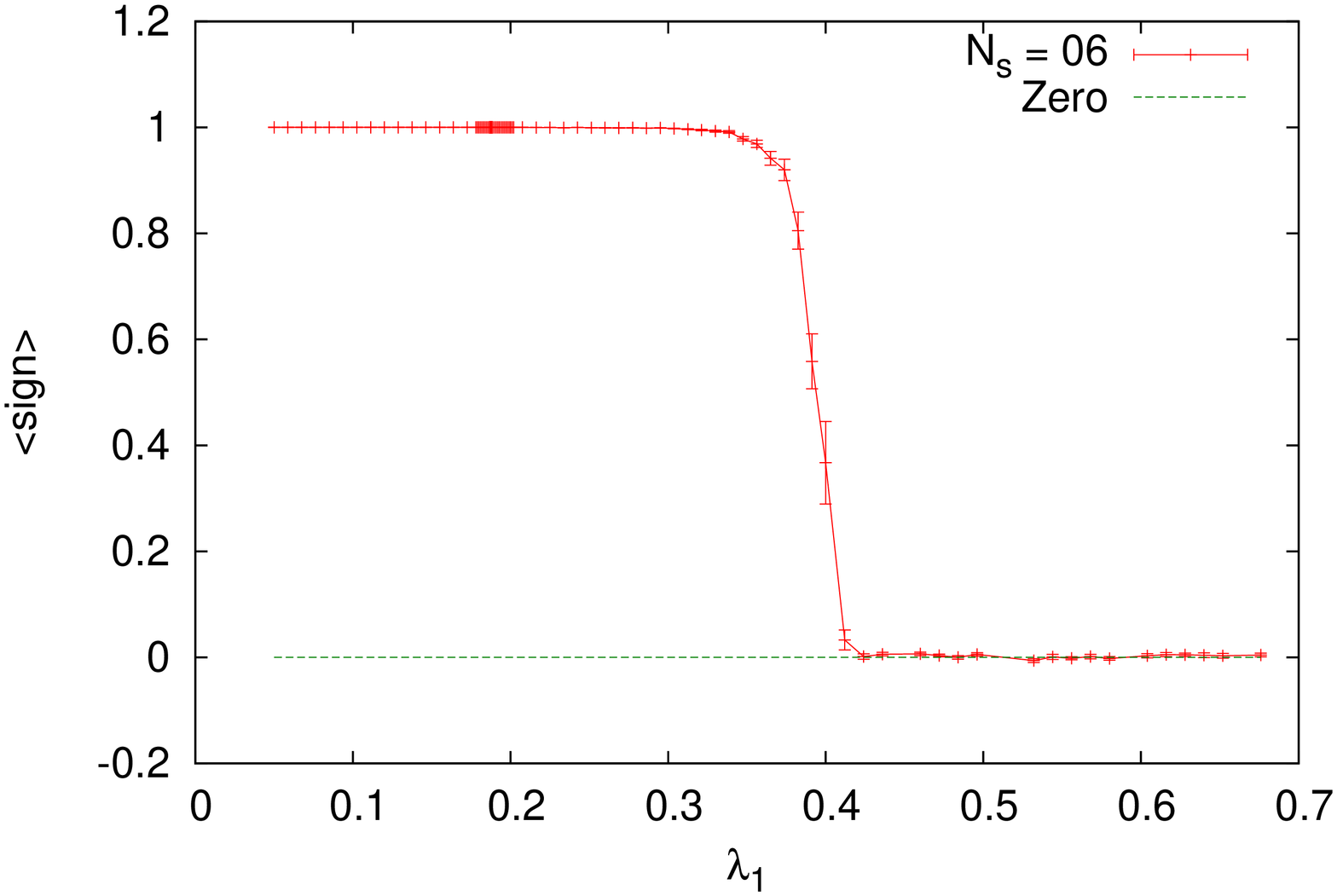} \hspace{0.6cm}
\includegraphics[height=5.5cm,angle=270]{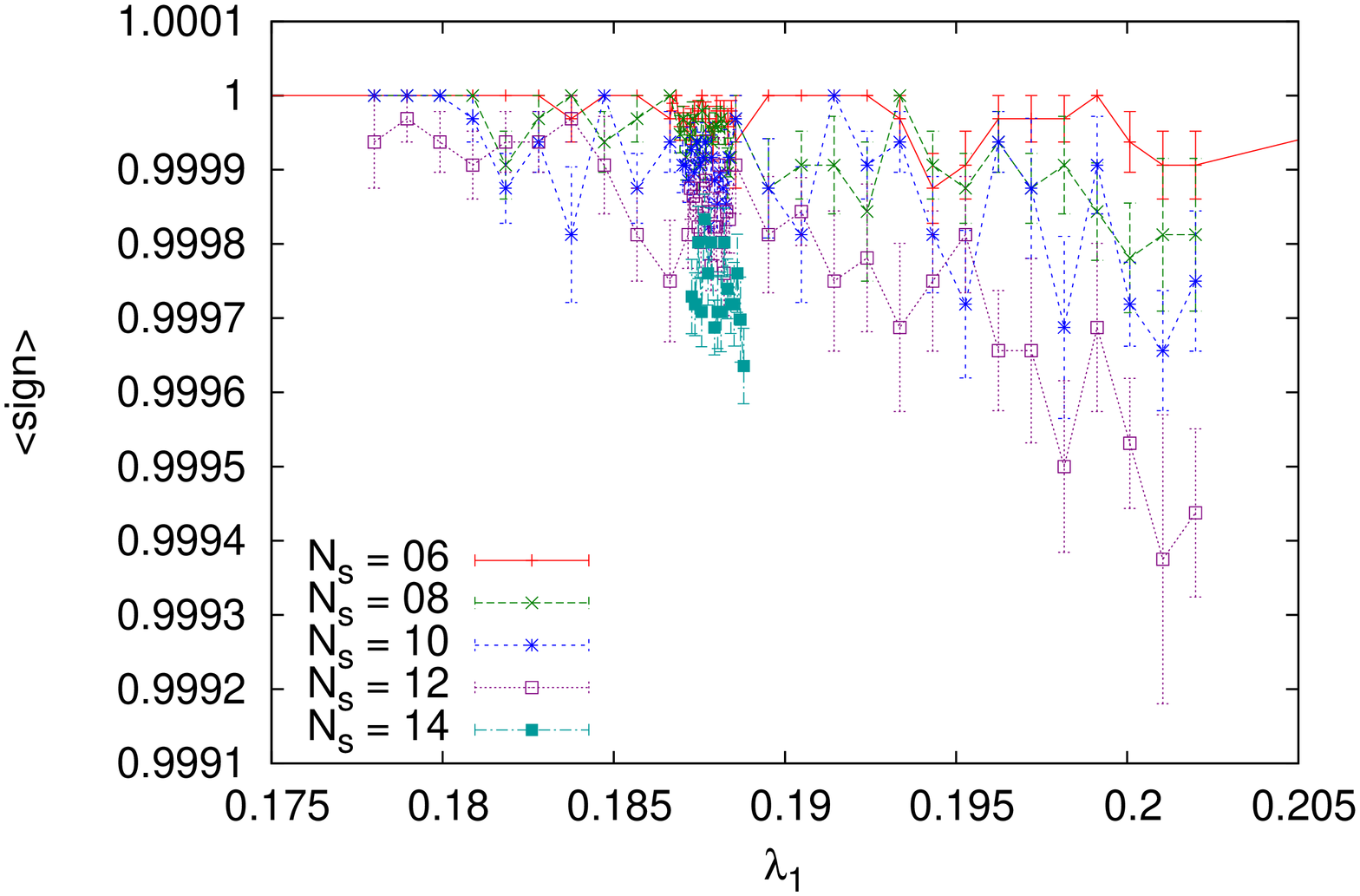} 
\caption{Average sign of $(1+2\lambda_1 \Real L_i L^*_j)$. Left: a wide range of $\lambda_1$ on a $N_s=6$ system. 
Right: zoom on the transition region with different system sizes.}
\label{fig:avgsign}
\end{center}
\end{figure}

In section (3.2) of \cite{Langelage:2010yr} we discussed the occurrence of a sign problem due to the 
factor $(1+2\lambda_1 \Real L_i L^*_j)$, and proposed a solution by approximating the corresponding 
exponentiated logarithm, $\exp\log (1+2\lambda_1 \Real L_i L^*_j)$, by a few terms of its series 
expansion up to order $M$, eq.~(3.6) in \cite{Langelage:2010yr}.
Fortunately, with the corrected measure the average sign of this factor remains $\geq 0.999$ for
the couplings of interest, $0<\lambda\lesssim \lambda_c\approx 0.188$, on all volumes simulated.
This is demonstrated in figure \ref{fig:avgsign}.
Hence, the corrected actions can be simulated exactly, such that section 3.2 of \cite{Langelage:2010yr} 
as well as approximations of the logarithm by polynomials of order $M$ are now obsolete.
In the sequel we list updated plots and tables replacing the previous ones.
We also report the corrected result for the one-coupling critical value (see figure \ref{fig9}):
\eq
	\lambda_{1,c}^{(1)} = 0.187885(30)\;\;.
\qe
Note that this value differs from the one obtained for the $M=1$-truncation of the theory,
$\lambda_{1,c}^{(1,M=1)}=0.13721(5)$ \cite{Wozar_et_al_2006}. The corrected critical lines for the 
two-coupling theories are  shown in figure \ref{fig12}. The $\beta_c(N_\tau)$
of the original theory are based on the maps eqs.~(2.21-2.23) in \cite{Langelage:2010yr}
and eq.~(5) in \cite{EQCD_Proceeding_2011}, with corrected values given in table 1.

\begin{figure}
\hspace*{-0.5cm}
\includegraphics[width=0.25\textwidth,angle=-90]{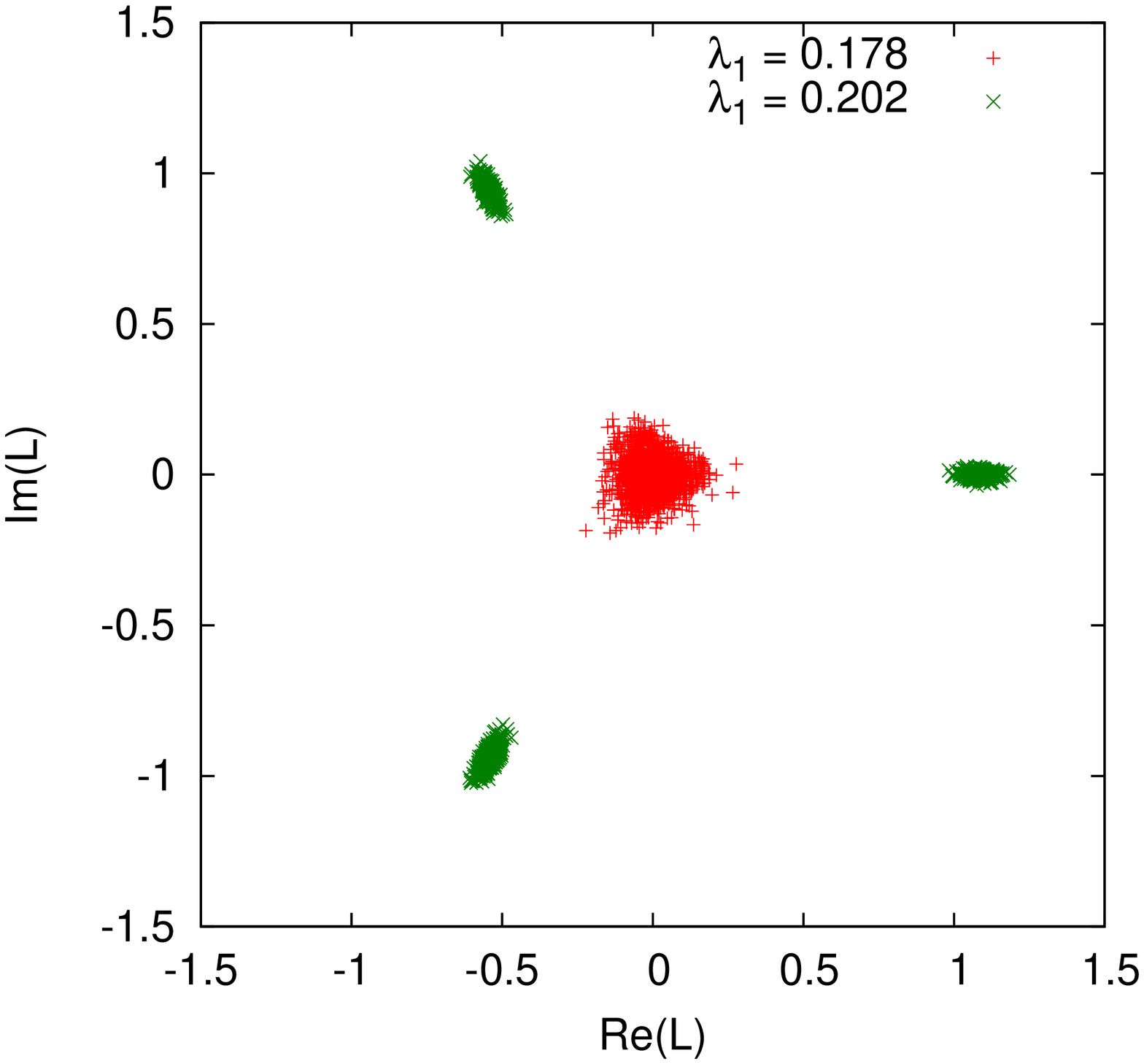}
\hspace*{-0.8cm}
\includegraphics[width=0.25\textwidth,angle=-90]{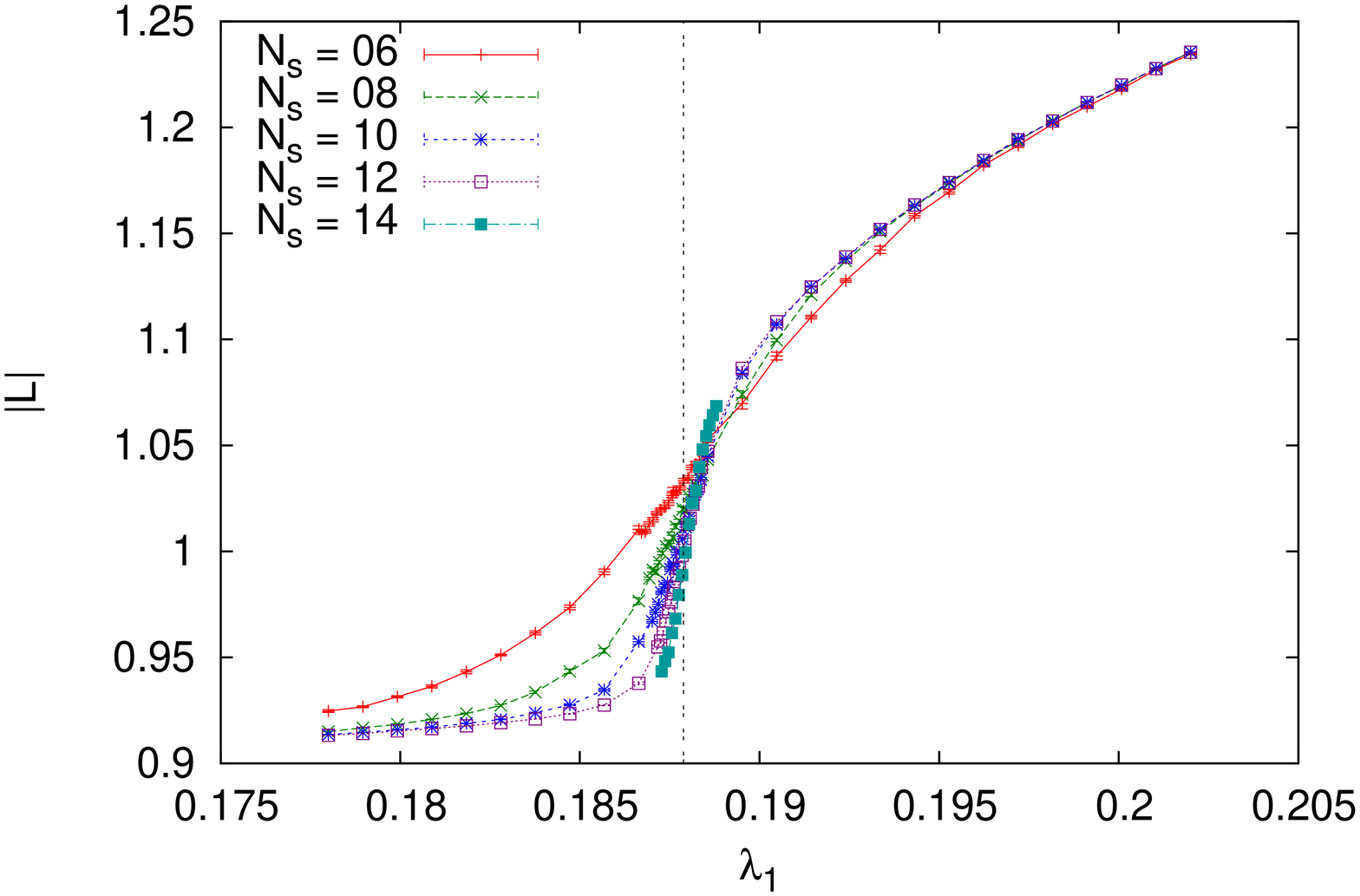}~\includegraphics[width=0.25\textwidth,angle=-90]{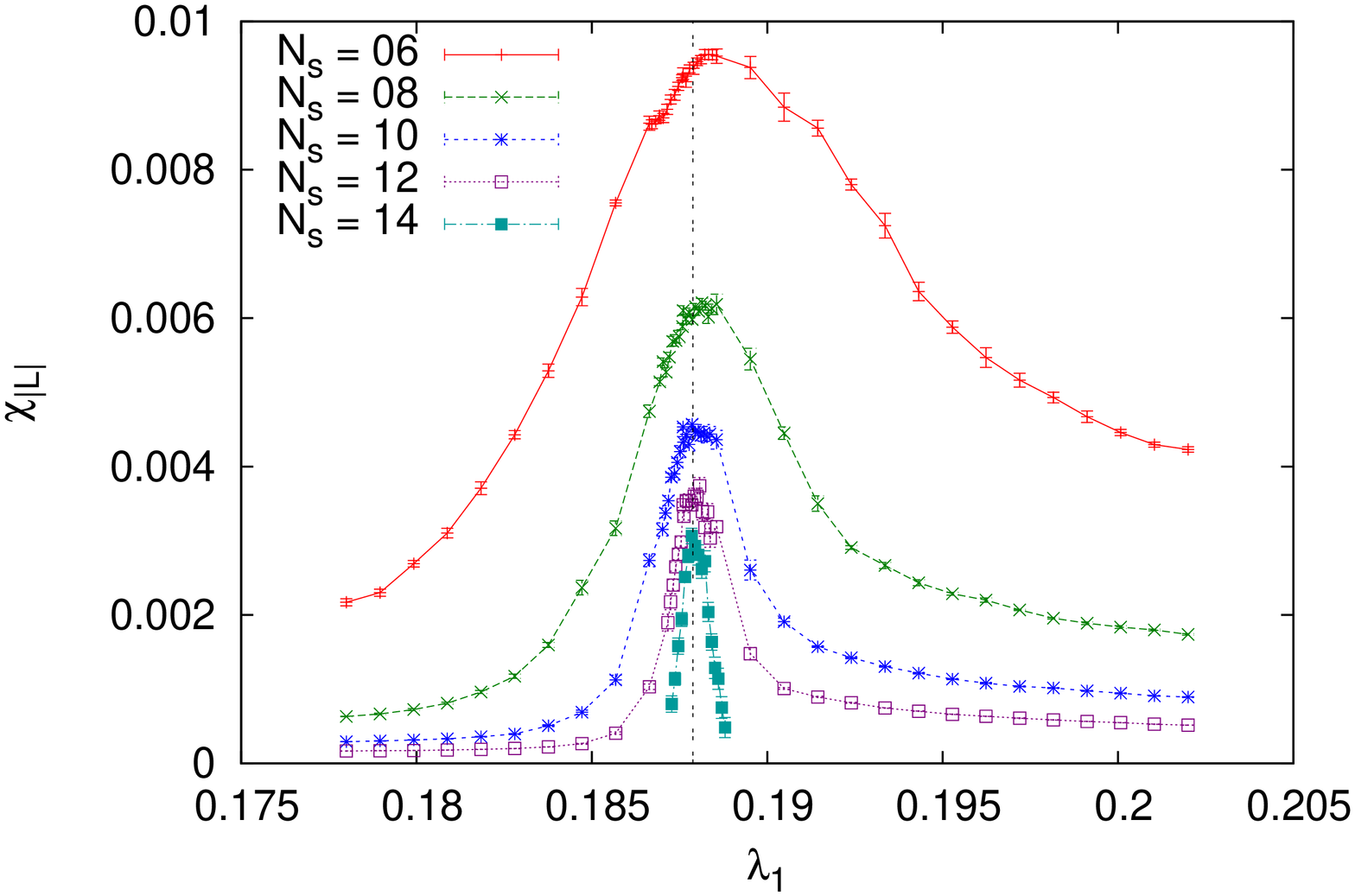}
\caption[]{Left: Distribution of $L$ for small and large $\lambda_1$. Middle, Right: Expectation
value of $|L|$ and its susceptibility. The vertical line marks the infinite-volume transition. Replaces figure 8
in \cite{Langelage:2010yr}. }
\label{fig:l_ERRATA}
\end{figure}
\begin{figure}
\begin{center}
\includegraphics[width=0.27\textwidth,angle=-90]{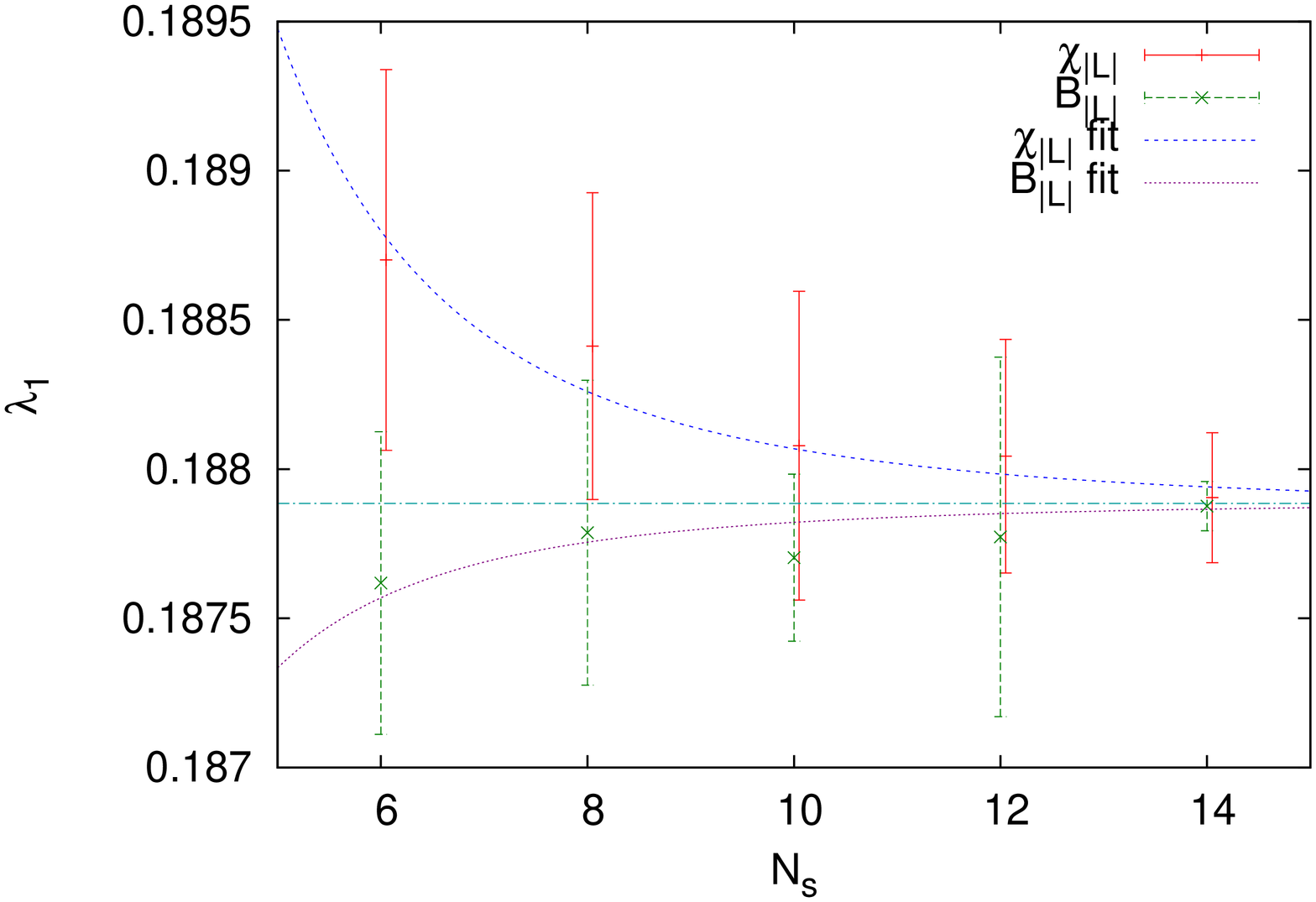}
%\hspace*{-0.8cm}
\includegraphics[width=0.27\textwidth,angle=-90]{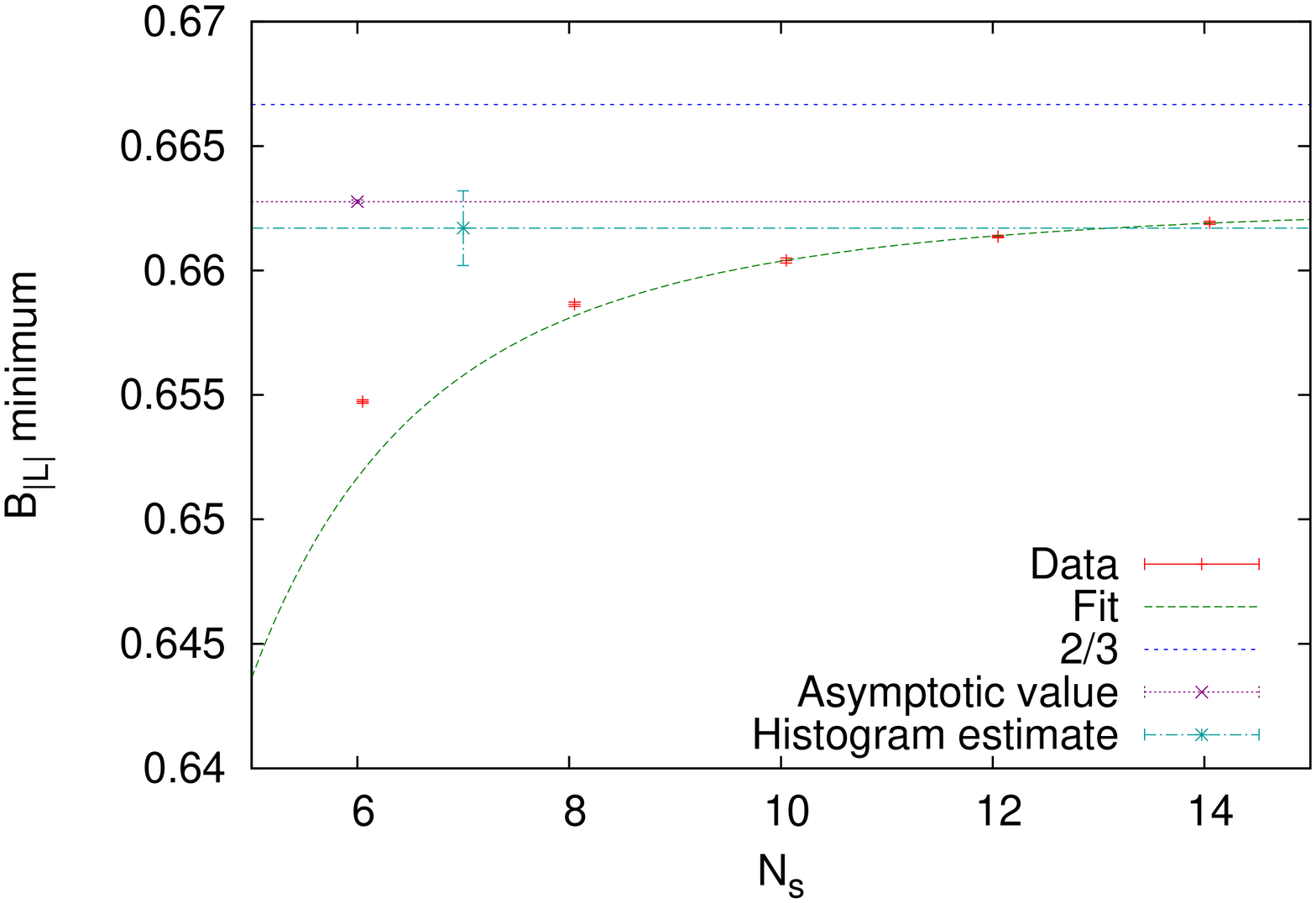}
\caption[]{Left: Position of the minimum of $B(|L|)$ and of the peak of $\chi(|L|)$ as 
defined in eqs.~(3.9) and (3.10) of \cite{Langelage:2010yr}. The curves 
are fits to eq.~(3.7) in \cite{Langelage:2010yr}, and the horizontal line is their common asymptotic value.
Replaces figure 9 in \cite{Langelage:2010yr}. }
\label{fig9}
\end{center}
\end{figure}
\begin{figure}
\begin{center}
\includegraphics[width=0.27\textwidth,angle=-90]{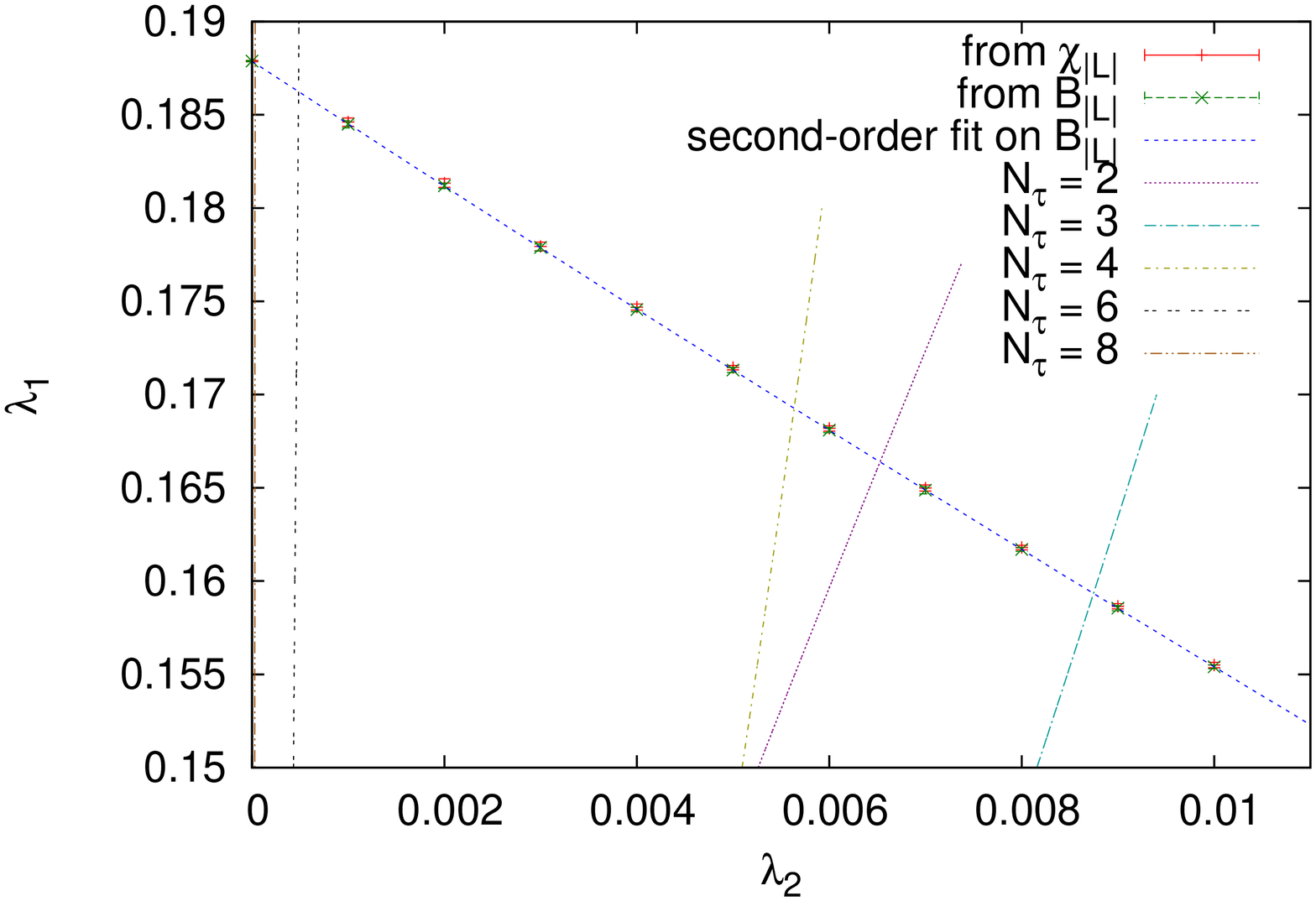}
%\hspace*{-0.8cm}
\includegraphics[width=0.27\textwidth,angle=-90]{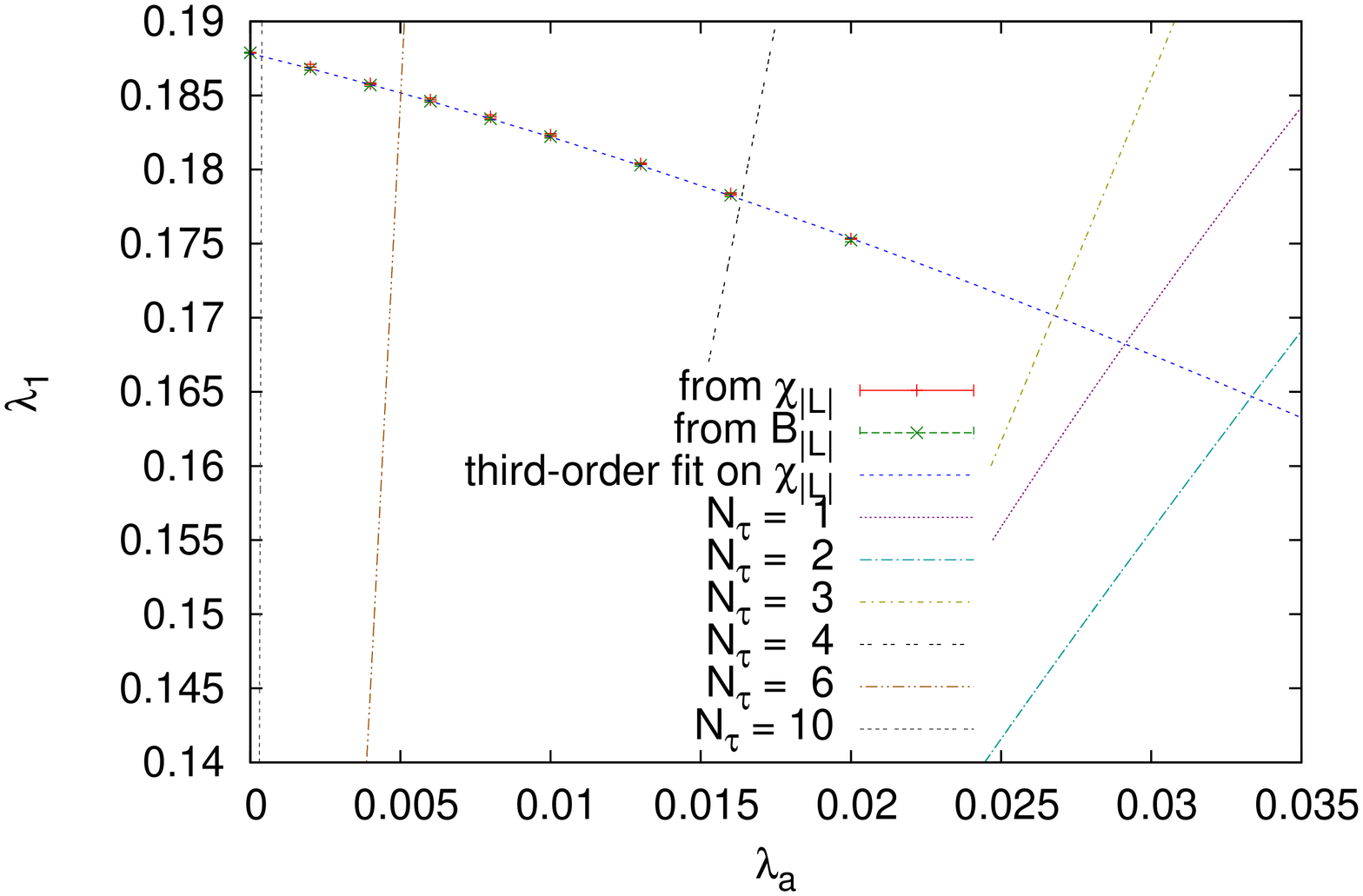}
\caption[]{Critical line in the two-coupling space, determined from $\chi(|L|)$.
	Intersection points give the parameters representing a 4d theory with fixed $N_\tau$.
	Left: $(\lambda_1,\lambda_2)$ Right: $(\lambda_1,\lambda_a)$. Replaces figure 14
in \cite{Langelage:2010yr}. }
\label{fig12}
\end{center}
\end{figure}

\begin{table}
\begin{center}
\begin{tabular}{|c||c|c|c||c|c|}
\hline
	$N_\tau$ &$\lambda_1$  & $(\lambda_1,\lambda_2)$ &
	$(\lambda_1,\lambda_a)$ & $\mbox{4d YM}$ & $\lambda_1,M=1$ \\
\hline
1   &  2.7828(4)   & --          & 2.529(6)   &   2.703(4) & 2.1150(7)  \\
2   &  5.1839(2)   & 5.0174(4)   & 5.003(5)   &  5.10(5)   & 4.7375(6)  \\
3   &  5.8488(1)   & 5.7333(3)   & 5.780(2)   &  5.55(1)   & 5.6226(3)  \\
4   &  6.09871(7)  & 6.0523(1)   & 6.0748(6)  &  5.6925(2) & 5.9552(2)  \\
6   &  6.32625(4)  &  6.32399(3) & 6.3225(1)  &  5.8941(5) & 6.2436(1)  \\
8   &  6.43045(3)  &  6.43033(2) & 6.42971(7) &  6.001(25) & 6.37245(7) \\
10  &  6.49010(2)  &  6.49008(2) & 6.48991(6) &  6.160(7)  & 6.44544(5) \\
12  &  6.52875(2)  & 6.52874(1)  & 6.52869(5) &  6.268(12) & 6.49244(4) \\
14  &  6.55584(2)  & 6.55583(1)  & 6.55580(4) &  6.383(10) & 6.52525(4) \\
16  &  6.57588(1)  & 6.57587(1)  & 6.57585(3) &  6.45(5)   & 6.54946(3) \\
\hline
\end{tabular}
\caption{Critical couplings $\beta_c$  for $SU(3)$ from different effective 
theories compared to simulations of the 4d theory \cite{fingberg_heller_karsch_1993, Kogut_et_al_1983}.
The $N_\tau=1$ Monte Carlo value is from our own simulation with standard 
Cabibbo-Marinari plaquette update.
In the last column $\beta_c$ is estimated from the $M=1$ truncation of the 
theory \cite{Wozar_et_al_2006}.
Replaces table 2 in \cite{Langelage:2010yr}.}
\label{tab:su3_betas_ERRATA}
\end{center}
\end{table}

\begin{figure}
\begin{center}
\includegraphics[width=0.27\textwidth,angle=-90]{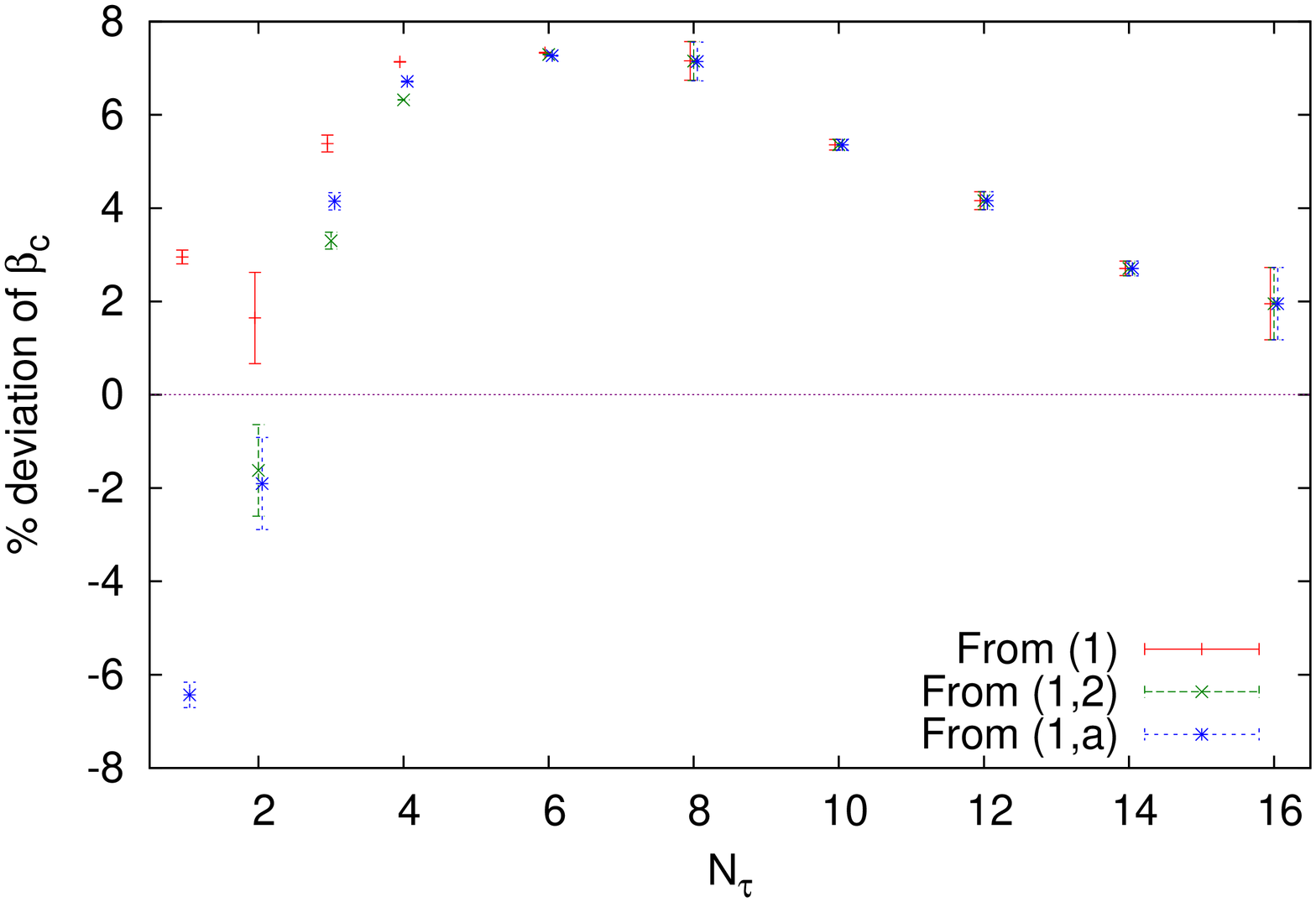}
%\hspace*{-0.8cm}
\includegraphics[width=0.27\textwidth,angle=-90]{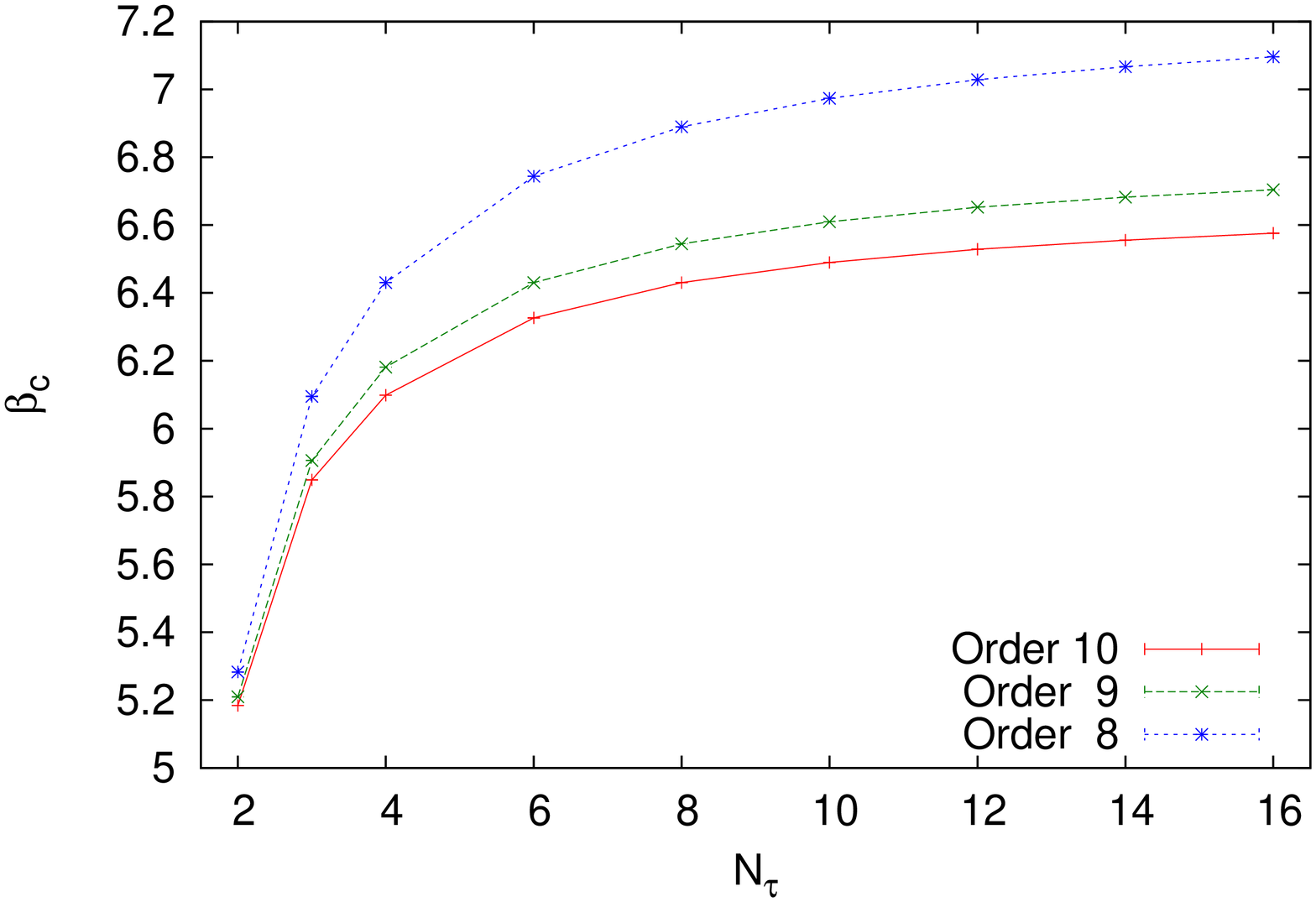}
\caption{Left: relative error of $\beta_c$ as estimated by the effective models compared to the Monte Carlo results 
(data are slightly shifted in the $x$-direction for ease of reading);
replaces figure 15 (right) in \cite{Langelage:2010yr}.
Right: assessment of the systematic errors in the strong-coupling series; replaces figure 16 (left) in \cite{Langelage:2010yr}.
}
\label{fig:finalplot}
\end{center}
\end{figure}

\section*{Acknowledgements}
We are indebted to Michael Fromm for pointing out the error in our original work.

\end{document}